\documentclass[usenatbib]{mn2e}
\newif\ifAMStwofonts

\usepackage{aas_macros}
\usepackage[footnotesize]{subfigure}
\usepackage{graphicx}

\title{The host galaxies of luminous quasars}

\author[D. J. E. Floyd et al.]  {David
       J. E. Floyd,$^{1}$\thanks{djef@roe.ac.uk} Marek
       J. Kukula,$^{1}$ James S. Dunlop,$^{1}$ Ross J. McLure,$^{1}$
       \newauthor Lance Miller,$^{2}$ Will J. Percival,$^{1}$ Stefi
       A. Baum$^{3}$ and Christopher P. O'Dea$^{3}$ \\ $^{1}$Institute
       for Astronomy, University of Edinburgh, Royal Observatory,
       Edinburgh EH9 3HJ, U.K. \\ $^{2}$Astrophysics, Department of
       Physics, Keble Road, Oxford, OX1 3RH, U.K. \\ $^{3}$Space
       Telescope Science Institute, 3700 San Martin Drive, Baltimore,
       MD 21218, U.S.A.} 

\date{  }

\pagerange{\pageref{firstpage}--\pageref{lastpage}} \pubyear{2003}

\begin{document}

\maketitle

\label{firstpage}

\begin{abstract}
  We present the results of a deep HST/WFPC2 imaging study of 17
  quasars at $z\simeq0.4$, designed to determine the properties of their
  host galaxies. The sample consists of quasars with absolute
  magnitudes in the range $-24\geq M_{V} \geq -28$, allowing us to
  investigate host galaxy properties across a decade in quasar
  luminosity, but at a single redshift.  Our previous imaging studies
  of AGN hosts have focussed primarily on quasars of moderate luminosity, 
  but the most powerful objects in the current sample have powers 
  comparable to the most luminous quasars found at high redshifts. 

  We find that the host galaxies of all the radio-loud quasars, and all 
  the radio-quiet quasars in our sample with nuclear luminosities 
  $M_V<-24$, are massive bulge-dominated galaxies, confirming and extending 
  the trends deduced from our previous studies. From the best-fitting 
  model host galaxies we have estimated spheroid and hence black-hole masses, 
  and the efficiency (with respect to the Eddington luminosity) with which 
  each quasar is emitting radiation. The largest inferred black-hole
  mass in our sample is $M_{BH} \simeq 3 \times 10^{9} {\rm
  M_{\odot}}$, comparable to the mass of the black-holes at the
  centres of M87 and Cygnus A. We find no evidence for super-Eddington
  accretion rates in even the most luminous objects ($0.05<L/L_{Edd}<1.0$).

  We investigate the role of scatter in the black-hole:spheroid
  mass relation in determining the ratio of quasar to host-galaxy
  luminosity, by generating simulated populations of
  quasars lying in hosts with a Schechter mass function. 
  Within the subsample of the highest-luminosity quasars, the observed
  variation in nuclear-host luminosity ratio is consistent with
  being the result of the scatter in the black-hole:spheroid
  relation.
  Quasars with high nuclear-host luminosity ratios 
  can be explained in terms of sub-Eddington accretion rates onto
  black-holes in the high-mass tail of the black-hole:spheroid
  relation.
  Our results imply that, owing to the Schechter function cutoff, 
  host mass should not continue to increase linearly with quasar
  luminosity, at the very highest luminosities.
  Any quasars more luminous than $M_V = -27$ should be found in
  massive elliptical hosts which at the present day would
  have $M_V \simeq -24.5$.
\end{abstract}


\setcounter{figure}{0}

\section{Introduction}
\label{sec-intro}

Thanks largely to the resolution advantage offered by the 
Hubble Space Telescope (HST) the last decade has seen
huge advances in our understanding of the host galaxies of the nearest
($z<0.3$) quasars (\citealt{bahcall+97}, 
\citealt*{hooper+97}, \citealt{boyce+98, mclure+99,
mcleod01}, \citealt*{hamilton+02}, \citealt{dunlop+03}). 
HST observations have demonstrated
that low-$z$ quasar hosts are luminous ($L > L^{\star}$) galaxies,
confirming the results of earlier ground-based studies. But the key
advantage of HST has been its ability to distinguish between disc and
elliptical morphologies, leading to the finding that all radio-loud
quasars (RLQs) and the majority of radio-quiet quasars (RQQs) reside in
massive bulge-dominated galaxies.

With recent improvements in the capabilities of both HST and 
ground-based telescopes, new studies are beginning to shed light on the
evolution of quasar hosts from high redshifts ($z>1$) down to the
present day (\citealt{lehnert+99b}, \citealt*{falomo+01},
\citealt{stocktonridgway01, ridgway+01, kukula+01, hutchings+02}).  At
the same time it has become 
increasingly clear from studies of inactive galaxies that black-hole
and galaxy formation and growth are intimately linked processes,
resulting in the now well-established correlation between black-hole
mass and the mass of the host galaxy's stellar bulge
\citep{magorrian+98, gebhardt+00, ferraresemerritt00}.

Most previous studies of quasar hosts have concentrated on quasars of
relatively low luminosity, largely because it is much easier to
disentangle the host and nuclear light in such objects. However, the
quasar population spans luminosities ranging from the (admittedly
somewhat arbitrary) transition from Seyfert galaxies at $M_{V}=-23$
through to the most luminous objects, with absolute magnitudes
$M_{V}\sim-30$; a factor of $\simeq1000$ in terms of luminosity. The
majority of quasars currently known at large redshifts belong to the
bright end of the luminosity function. This is due to the
degeneracy between redshift and luminosity in flux-limited samples, a
situation which is beginning to be rectified as modern deep surveys
detect low-luminosity, high-$z$ quasars in increasing
numbers. However, to understand the behaviour of the most massive
galaxies and their black-holes in the cosmologically-interesting high
redshift regime ($z>2$) will inevitably require the study of the most
luminous quasars at these redshifts.

The aim of the current study is to help break the degeneracy between quasar
luminosity and redshift by studying a sample of quasars at a single
redshift that spans an appreciable fraction of the quasar luminosity
range (see Fig.1). The lowest redshift at which this can be attempted is
$z\simeq0.4$, since the cosmological volume enclosed at lower redshift
is too small to contain examples of the rare, most luminous quasars, 
with $M_{V} < -27$.
Not only does the design of this study 
allow us to explore the relation between quasar
luminosity and host-galaxy properties, but the most
luminous objects in this programme can also provide a low-redshift
baseline against which to compare the hosts of luminous high-$z$
quasars in future studies.

The key technical difficulty in this work is, of course, the
disentanglement of the point spread function (PSF) produced by the
bright unresolved nucleus, from the emission produced by the host
galaxy. In this study, as in~\citet{dunlop+03}, we have used 2D
modelling to perform this decomposition. 

The paper is structured as follows. In Section 2 we describe the
quasar sample along with our selection criteria. Section 3 details our
observing strategy and Section 4 describes the reduction of the HST
images. The 2-D modelling procedure used to analyse the images and
extract information about the hosts is described in Section 5. Section
6 summarises the results produced by this image modelling, with the
images themselves being located in an Appendix, along with
detailed information about each object. In Section 7 we discuss our
principal results and their implications for our understanding 
of the quasar phenomenon. Finally, our conclusions are summarised in
Section 8. 

For ease of comparison with our previous work we adopt an Einstein-de
Sitter universe with $H_{0}=50$ km s$^{-1}$ Mpc$^{-1}$.


\section{The quasar sample}
\label{sec-samp}
\begin{table*}
\caption{Quasars in the current study. J2000 co-ordinates were
obtained from the Digitised Sky Survey plates maintained by the Space
Telescope Science Institute. Redshifts and apparent $V$ magnitudes are
from the quasar catalogue of \citet{VCV2000}. For consistency
we use a B1950 IAU format to refer to the quasars in this paper; the
name under which each object appears in the HST Archive is given in
the final column and additional names are given in the description of
each object in the Appendix. The low-luminosity subsample was observed in
HST Cycle 7, using the WF2 chip and the F814W filter. The
high-luminosity subsample was observed in Cycle 9 using the WF3 chip,
and the slightly narrower F791W filter. This latter observing run
included one additional object, 1404$-$049, an inactive spiral galaxy
at redshift 0.04 which had been misclassified as a quasar. This object
has been omitted from the analysis presented in this paper.}
\begin{tabular}{lccccccrr}
\hline
\hline
Name (B1950)& Type&$z$ & RA (J2000)& Dec (J2000)&$V$&$M_{V}$&Observing date&HST archive name\\
\hline
\multicolumn{9}{c}{Low-luminosity subsample (Cycle 7; F814W)}\\
1237$-$040 &RQQ&0.371 &$12:39:39.0$&$-04:16:38.5$&$16.96$&$-24.77$&Feb 12 1999&1239$-$041 \\
1313$-$014 &RQQ&0.406 &$13:16:09.0$&$-01:54:54.1$&$17.54$&$-24.39$&Feb 01 1999 &Q1313$-$0138\\
1258$-$015 &RQQ&0.410 &$12:58:13.9$&$-02:00:09.3$&$17.53$&$-24.42$&Feb 19 1999&1258$-$015 \\
1357$-$024 &RQQ&0.418 &$14:00:06.4$&$-02:42:22.6$&$17.43$&$-24.56$&Mar 01 2000&1400$-$024 \\
1254$+$021 &RQQ&0.421 &$12:57:05.9$&$+01:49:46.8$&$17.14$&$-24.86$&Feb 11 1999&1257$+$015 \\
1150$+$497 &RLQ&0.334 &$11:53:22.3$&$+49:30:21.4$&$17.10$&$-24.40$&Nov 25 2000&LB2136      \\
1233$-$240 &RLQ&0.355 &$12:35:39.6$&$-25:11:31.1$&$17.18$&$-24.46$&Aug 09 1997&PKS1233$-$24\\
0110$+$297 &RLQ&0.363 &$01:13:22.0$&$+29:58:58.8$&$17.00$&$-24.68$&Feb 02 1999&B2$-$0110$+$29 \\
0812$+$020 &RLQ&0.402 &$08:15:21.3$&$+01:55:44.8$&$17.10$&$-24.80$&Feb 18 1999&PKS0812$+$02\\
1058$+$110 &RLQ&0.423 &$11:00:49.3$&$+10:47:00.3$&$17.10$&$-24.90$&Apr 02 1999&AO1058$+$11 \\
GRW$+$70D5824& Star & -- &$13:38:58.0$&$+70:16:32.1$&$12.77$& -- &Feb 25 2000&PSF-STAR\\ 
\hline
\multicolumn{9}{c}{High-luminosity subsample (Cycle 9; F791W)}\\
0624$+$691 &RQQ&0.370 &$06:30:08.6$&$+69:05:40.0$&14.20&$-27.73$&May 25 2000&HS0624$+$6907\\
1821$+$643 &RQQ&0.297 &$18:22:02.8$&$+64:20:05.3$&14.10&$-27.31$&Aug 17 2000&E1821$+$643\\
1252$+$020 &RQQ&0.345 &$12:55:22.5$&$+01:43:46.2$&15.48&$-26.29$&May 16 2000&EQS$-$B1252$+$020\\
1001$+$291 &RQQ&0.330 &$10:04:06.1$&$+28:55:19.2$&15.50&$-26.16$&May 21 2000&TON0028 \\
1230$+$097 &RQQ&0.415 &$12:33:28.8$&$+09:31:04.9$&16.15&$-26.05$&Jul 5  2000&1230$+$097\\
0031$-$707 &RLQ&0.363 &$00:33:57.2$&$-70:25:24.5$&15.50&$-26.39$&Aug 7  2000&MC4\\
1208$+$322 &RLQ&0.388 &$12:10:39.8$&$+31:56:26.0$&16.00&$-26.04$&May 7  2000&B2$-$1208$+$32\\
GRW$+$70D5824&Star&-- &$13:38:54.1$&$+70:16:21.2$&12.77& -- &May 1  2000&PSF-STAR\\
\hline
\hline
\end{tabular}
\end{table*}

The sample was selected from the quasar catalogue of \citet{VCV1993}
and comprises two subsamples, both confined to the 
redshift range $0.29 < z < 0.43$
(Table~1). The first, `low-luminosity' subsample consists of 
five radio-loud and five radio-quiet quasars (RLQs \&
RQQs) with absolute magnitudes $-24 > M_{V} > -25$. All of the RLQs
have 5\,GHz radio luminosities $> 10^{24}\,$W~Hz$^{-1}$sr$^{-1}$ and steep
radio spectra to ensure that their radio luminosities have not 
been significantly boosted by
relativistic beaming. The RQQs have all been surveyed in the radio at
sufficient depth to ensure that their 5\,GHz luminosities are indeed
$<10^{24}\,$W~Hz$^{-1}$sr$^{-1}$.  
The second, `high-luminosity' subsample consists of all known quasars
in this redshift range with absolute magnitudes $M_{V} < -26$, and
includes 2 quasars with $M_{V} \simeq -28$.
These two samples allow us to explore an orthogonal direction in the
optical luminosity - redshift plane, in contrast to our previous HST
studies of quasar hosts \citep{mclure+99, kukula+01, dunlop+03} which
concentrated on quasars of comparably moderate luminosity ($M_{V} >
-25$), but spanning a wide range in redshift out to $z\simeq 2$
(Fig.1).

\begin{figure}
\label{fig-samp}
\centering
\includegraphics[width=80mm]{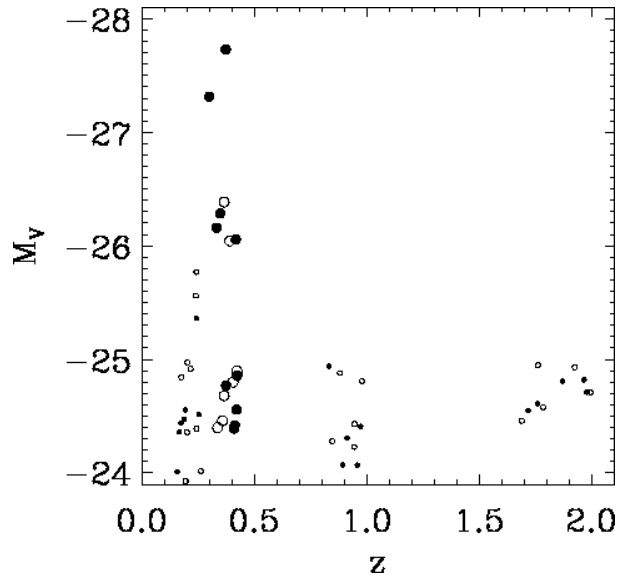}
\caption{Absolute magnitude versus redshift for quasars observed to
date in our HST host-galaxy imaging programmes. Filled circles
represent radio-quiet quasars (RQQs) and open circles radio-loud
quasars (RLQs). Our earlier work (small symbols) concentrated mainly
on quasars of moderate luminosity (typically $M_{V}>-25$) in
three redshift regimes ($z \simeq 0.2, 1$ \& 2), allowing us to probe
the evolution of the host galaxies over a large fraction of cosmic
history \citep{mclure+99,kukula+01,dunlop+03}. 
The current study (large symbols) is designed to explore an
orthogonal direction in the $M_{V} - z$ plane, by sampling a large
range of quasar luminosities at a single redshift, $z\simeq 0.4$. This
is the lowest redshift at which one can find very luminous quasars (those with
$M_{V} < -27$), comparable to the most luminous quasars
in the high-redshift universe.}
\end{figure}

\begin{table}
  \caption{$V$-band magnitudes and 5 GHz radio flux densities for the
    quasars in our sample. At the time of writing the majority of the RQQs
    have only upper limits to their radio fluxes, but these are sufficient
    to place them safely below the accepted radio-loud/radio-quiet
    boundary of $L_{5GHz} < 10^{-24}$ W~Hz$^{-1}$ sr$^{-1}$. The final
    column gives the source of the radio data (VCV=\citet{VCV2000};
    G+99=\citet{goldschmidt+99}; B+96=\citet{blundell+96}. Upper limits
    from the NVSS were converted to 5 GHz by assuming a radio spectral
    index $\alpha=0.5$ ($f_{\nu} \propto \nu^{-0.5}$).}
  \centering
  \begin{tabular}{ccccrrl}
    \hline
    \hline
    Source  &Class&$z$  &$V$  &$S_{5GHz}$&$L_{5GHz}$& Ref.\\
    &     &     &     & (mJy)    &          &          \\
    \hline
    \multicolumn{7}{c}{Low-luminosity}\\
    1237$-$040&RQQ&0.371&16.96&  $<0.3$& $<22.14$ &G+99\\
    1313$-$014&RQQ&0.406&17.54&  $<0.3$& $<22.21$ &G+99\\
    1258$-$015&RQQ&0.410&17.53&  $<0.2$& $<22.05$ &G+99\\
    1357$-$024&RQQ&0.418&17.43&  $<0.3$& $<22.24$ &G+99\\
    1254$+$021&RQQ&0.421&17.14&  $<0.3$& $<22.25$ &G+99\\
    1150$+$497&RLQ&0.334&17.10&   717.0&  25.62 &VCV\\
    1233$-$240&RLQ&0.355&17.18&   670.0&  25.45 &VCV\\
    0110$+$297&RLQ&0.363&17.00&   340.0&  25.17 &VCV\\
    0812$+$020&RLQ&0.402&17.10&   845.0&  25.62 &VCV\\
    1058$+$110&RLQ&0.420&17.10&   225.0&  25.12 &VCV\\
    \hline
    \multicolumn{7}{c}{High-luminosity}\\
    0624$+$691 &RQQ&0.370&14.20& $<1.2$& $<22.75$ &NVSS\\
    1821$+$643 &RQQ&0.297&14.10&    8.6&  23.58   &B+96\\
    1252$+$020 &RQQ&0.345&15.48&    0.8&  22.50   &G+99\\
    1001$+$291 &RQQ&0.330&15.50& $<1.2$& $<22.65$ &NVSS\\
    1230$+$097 &RQQ&0.415&16.15& $<1.2$& $<22.85$ &NVSS\\
    0031$-$707 &RLQ&0.363&15.50&   95.0&  24.62   &VCV\\
    1208$+$322 &RLQ&0.388&16.00&   91.0&  24.90   &VCV\\
    \hline
    \hline
  \end{tabular}
\end{table}


\section{Observing strategy}
All of our previous HST observations of quasar hosts were carefully
designed to maximize the chances of successfully separating the
starlight of the host from the PSF of the
central quasar. We used the same observing strategy for the current
observations and, since some of the quasars in our new sample are
significantly more luminous than those in the earlier programmes,
these precautions assume even greater importance.

Observing dates for each object are listed in Table~1, along with the
name under which the dataset is listed in the HST Archive.  The
observations were carried out in two different HST observing cycles,
although in practice the dates overlap. Observations of the
low-luminosity subsample were carried out in Cycle 7 whilst the
high-luminosity subsample was observed as part of Cycle 9.

\subsection{Choice of filter}
As in our previous programmes we selected filters to correspond to
$V$-band in the quasar's restframe. This ensures that our images
sample the object's restframe spectrum longwards of the 4000\,\AA\,break, 
where the starlight from the host is relatively bright, whilst
avoiding strong emission lines such as H$\alpha$ and 
[O{\sc iii}]$\lambda 5007$. Despite being directly associated with the quasar
activity, ionised emission-line regions can extend over several
kiloparsec. By excluding such emission from the images, we obtain a
cleaner picture of the distribution of starlight in the hosts.

For the low-luminosity subsample we used the F814W `broad $I$' filter
which corresponds closely to the standard Cousins $I$-band. The
high-luminosity subsample spans 
a slightly broader range of redshifts and in
order to avoid contamination of the images by emission lines we used the
slightly narrower F791W filter. 

\subsection{Choice of detector}
Observations were made with the HST's Wide Field \& Planetary Camera 2
(WFPC2).  We opted to use the WF chips ($800 \times 800$ pixels, with
a scale of 100 mas pixel$^{-1}$) since their relatively large pixels offer
greater sensitivity to low surface brightness emission.  Targets were
centred on one of the three WF chips, the exact choice depending on
which of the three had performed best over the period immediately
prior to the Phase 2 proposal deadline.

\subsection{Exposure times}
High dynamic range is imperative in a study of this kind in which it
is necessary to accurately characterise both the central core of the
quasar as well as the faint outer wings of the PSF and the underlying
host. The approach used is the same as that tried and tested in our
previous works (e.g. McLure et al. 1999, Kukula et al. 2001, Dunlop et
al. 2003). We took several exposures of increasing length, with exposure
times carefully scaled so that no image would saturate beyond the
radius out to which the PSF could be followed in the previous, shorter
exposure. The series of exposures were then spliced together in
annuli to construct an unsaturated image of the requisite depth
(the pointing stability of HST between successive exposures using the
FGS fine tracking mode is $\simeq0.003$ arcsec). 

For the `low-luminosity' subsample a single orbit of HST time was
sufficient for each object, with exposures of 5, 26 and $3\times600$
seconds.  For the more luminous quasars in the second subsample we
required some shorter exposures to avoid saturation of the central
pixels, as well as more long exposures to provide greater depth, since
the wings of the quasar PSF encroach further out into the surrounding
galaxy.  Here we devoted two orbits to each object, with exposures of
2, 26, $2\times100$ and $3\times600$ seconds in the first orbit and
$3\times700$ seconds in the second.

\subsection{PSF determination}
Our previous host-galaxy studies have emphasized the importance of
characterising the instrumental PSF over a
large range in angular radius in order to accurately separate the
contributions of host galaxy and active nucleus. The structure of the
HST PSF is quite variable, especially at large radii, and depends on
its position in the instrument field of view, the SED of the target
source and the timing of the observations.

We therefore devoted two orbits of our HST time allocation to
obtaining deep, unsaturated stellar PSFs through both the F814W and
F791W filters. The star used was GRW$+$70D5824 ($V=12.77$), the same
white dwarf used in our previous quasar host-galaxy studies with HST
\citep{mclure+99,dunlop+03}.  This star is an optical standard for
WFPC2, and so very accurate photometry is readily available. Its DA3
spectral type and neutral colour ($B-V=0.09$) mimics well the typical
quasar SED at the redshift of our sample.  In addition, as
there are no comparably bright stars within 30 arcsec, we can be sure
that our stellar PSF is not contaminated by light from nearby
objects. All observations were performed using the same region of the
WF chip, within the central 50$\times$50 pixel region.

For the PSF observations we 
used an observing strategy similar to that for the quasars in order
to obtain final images with high dynamic range. A series of exposures
were carried out with durations of 0.23, 2.0, 26.0 and 160.0 seconds.
Although the increasing exposure times lead to increasing saturation
in the core, they never become saturated outside the radius at which
the signal-to-noise of the preceding (shorter) exposure has become
unsatisfactory. A 2-point dither pattern was also adopted in order to
achieve better sampling (the $I$-band PSF is under-sampled by the
$0.1$-arcsec pixels of the WF chips). Such a strategy is not employed
for the quasar observations, since the read noise introduced by the
additional observations results in an unacceptable decrease in
signal-to-noise ratio.


\section{Data reduction}
\label{sec-datared}

All the images were passed through the standard WFPC2 pipeline
software which performs many of the initial image processing and
calibration steps such as dark and bias frame subtraction, along with flat
fielding. We carried out three additional reduction steps prior to
analysing the images: cosmic-ray decontamination, sky subtraction, and
reconstruction of the saturated core regions of both quasar and
stellar PSFs by splicing in images with shorter exposures.

\subsection{Removal of cosmic rays and bad pixels} 
Cosmic rays were removed using {\sc CRREJ}, an iterative sigma-clipping
algorithm in {\sc iraf}, which rejects high pixels from sets of
exposures of the same field.  The result is a single deep image (with
integration time equal to the sum of its parts).  This technique
however cannot remove the numerous persistent bad or ``hot'' pixels
that appear in the WFPC2 chips. These are listed in a tabular form
on the WFPC2 web page and can be incorporated into a mask for modelling
purposes (see Section 5).

\subsection{Background subtraction}
In order to remove a smoothly varying sky
background from each image we adopted the following procedure. First, the
mean and standard deviation of the background in each image were
estimated from a subset of pixels excluding obvious sources. An
estimate of the sky distribution was then created by fitting a 2nd
order polynomial to each image, having replaced 1$\sigma$ deviations
away from the mean background level by the mean value. This smooth sky
map was subtracted from each original image to give a final
sky-subtracted frame. The background on WFPC2 images is extremely
flat, and higher order approximations found to be unnecessary.

\subsection{Building deep, unsaturated quasar images}
After cosmic ray removal from the deep images, the central regions
from the shorter exposures were cut out, multiplied by a factor determined
by annular photometry, and substituted into the saturated cores. All
steps were performed using standard data-reduction tasks within
{\sc iraf}. For the brightest quasars in our sample, in which the saturated
region was large, we spliced together three images in this way to
create the final image.


\section{Image analysis and modelling}
\label{sec-anal}
The techniques and software used to perform the 2D modelling have
been discussed previously in detail \citep*{mclure+00}, and in this
Section we only briefly reconsider the procedure, highlighting the
modifications required to cope with the extreme luminosities of some of
the quasars in the current study.

\subsection{Quasar models} 
Each model quasar was constructed by combining a model host galaxy
with a central delta function to represent the nuclear point source. We
model the host galaxy surface flux using the S\'{e}rsic equation
\citep{sersic68}: 
\begin{equation}
\label{eqn-sersic}
\Sigma(r)=\Sigma_{0} \exp\left[ -\left(\frac{r}{R_{0}}\right)^{\beta} 
\right]
\end{equation}
The surface brightness $\Sigma(r)$ describes an azimuthally-symmetric 
distribution, which is projected on to a generalised elliptical
coordinate system to allow for different eccentricities and
orientations of the host galaxy. 
$R_{0}$ is the characteristic scale length, $\Sigma_{0}$ is the
central surface brightness, and $\beta$ describes the overall
shape of the profile. We followed the procedure of \citet{mclure+99} and
\citet{dunlop+03} in fitting {\it first} to {\it a priori}
elliptical ($\beta=0.25$) and disc ($\beta=1.0$) models, and then to a
more general form in which $\beta$ is a free parameter. 
Throughout this work we use the term ``scale length'' to refer to the
half-light radius of the host galaxy models; i.e. the radius within
which half of the integrated light is contained. For elliptical
galaxy models, this half-light radius is equivalent to
the often quoted effective radius (or $R_{e}$), while for the disc or
spiral models ($\beta=1.0$) the half-light radius and the exponential
radius are related by $R_{1/2}=1.68 R_{0}$.

Care must be taken in translation from the continuous distribution
described by equation 1, to the discrete regime of CCD pixels. 
Simply basing the surface brightness of each pixel upon the radius of
the centre of that pixel is inaccurate, particularly toward the
centre, where $\Sigma(r)$ varies non-linearly across the angular width of a 
pixel. To account for this, galaxy models were calculated at a much
higher resolution, and then rebinned to WF chip resolution for
comparison with the data. Each pixel value depends upon at least 9
calculations of surface brightness, and up to 676 calculations within the
central 0.5 arcsec.  Flux is then added to the central pixel of the model
galaxy, to represent the unresolved nuclear component.

A given model quasar is thus specified uniquely by a point in the
6-dimensional parameter space $\{ \bmath{X} \} = \{ L_{N}, \Sigma_{0}, r_{0},
\beta, \Theta, \epsilon \}$:
\begin{itemize}
\item $L_{N}=$ luminosity of the nucleus
\item $\Sigma_{0}=$ central surface brightness of the host galaxy
\item $r_{0}=$ characteristic scale length of the host galaxy
\item $\Theta=$ position angle of the host galaxy
\item $\epsilon=\frac{a}{b}=$ axial ratio of the host galaxy
\item $\beta$ controlling the shape of the profile
\end{itemize}
The result is an idealised, seeing- and diffraction-free image of a quasar,
which can then be 
convolved with the appropriate PSF to produce a simulated HST
observation.  

\subsection{Modelling the PSF}
HST/WFPC2 offers both extremely deep imaging and an extremely well
characterised PSF. However, the PSF is under-sampled on the WF chips
and care was therefore taken to match the central regions of the PSF to
each quasar image. The sub-pixel centring of each quasar image was
found using the {\sc centroid} routine in {\sc iraf} to characterise the
distribution of light in the central region. Accurate oversampled
models of the central regions of the PSF were then generated using the 
{\sc tinytim} software \citep{tinytim} and were re-sampled using the correct central
position to provide an accurate model of the central few pixels of
each quasar image.

The {\sc tinytim} calculation depends upon optical path differences
within the Optical Telescope Assembly, and can be performed for any
sampling rate, or position within the WFPC2 field-of-view. Using
21-times oversampled (with respect to a WF pixel) PSFs rebinned to the
WF chip resolution, we matched the sub-pixel position of the centre of
the PSF to each quasar through a 2D minimum $\chi^2$ grid search. The
best-fit {\sc tinytim} model was then scaled up (by annular photometry) and
spliced into the centre of the deep stellar PSF image. This PSF could
then be convolved with the model galaxy plus nuclear component and the
result compared to the real quasar image.

\subsection{Pixel Error Analysis}
The error allocation for such a technique must be done carefully if
the $\chi^{2}$ figure of merit is to have real meaning.
Inaccurate error weighting may lead to the dominance of one region
over another in the fitting, and hence to biased results. Pixel values
were assumed to be independent and to obey Poisson statistics. 
We used a combination of Poisson and sampling errors.

\subsubsection{Poisson Noise Error}
The minimum possible noise in a given pixel is the combination of
Poisson error due to photon shot noise, dark current, and the noise
introduced by CCD read-out. However, this
simple error calculation severely underestimates the effective error in the
central region of our quasar images, due to sub-critical sampling of the
steeply rising PSF by the WF pixels.

\subsubsection{Sampling Errors}
For each quasar image we constructed a ``PSF residual'' frame by scaling 
up the PSF to match the quasar in the central pixel and subtracting. 
The resulting frame can be used to quantify the extent to which 
the re-sampled {\sc tinytim}+stellar PSF constructed for that particular 
quasar image has succeeded in mimicking the central, pixelised flux 
distribution. Specifically, we compute the inferred remaining sampling
error as a function of radius by calculating the variance of the 
distribution of pixel values in one-pixel wide annuli in this residual
image. This variance ($\sigma^{2}$) value is
then assigned to each pixel in a given annulus. This is done from the
centre out, until the value of this sampling error has fallen
(as expected at some radius, given adequate knowledge of the PSF) to
the level of the mean Poisson noise error discussed
above. We call this radius the ``sampling error radius'', $R_{samp}$. 
We could choose to use this approach across the whole image,
since it falls to the same level as the average 
Poisson noise by a radius of typically 
$\simeq 1$ arcsec from the centre of the quasar image.
However, it is clearly preferable, where possible, 
to assign each pixel an error based upon its own individual noise properties,
rather than a blanket error for an entire annulus.
The contribution of genuine galaxy residuals to this error calculation
are negligible.

For the two most luminous objects in this study, E1821 and HS0624, 
the sampling error in the PSF residual image remained above the mean Poisson 
noise at radii much larger than 1 arcsec. This is because, in these two cases
the central quasar is so bright that the image actually 
contains more information 
on the detailed structure of the PSF at large radii than does our deepest 
image of the PSF star. Consequently, for these two objects, the errors in our 
knowledge of the PSF at radii of several arcsec become important, and we had 
to enhance the adopted errors in the image at large radii (by typically
$\sqrt 2$; in practice we adopted an average of the `sampling' and Poisson 
errors at large radius) to achieve an acceptable model fit with a flat 
distribution of values in the final $\chi^2$ image produced by the 
model-fitting procedure.

\subsubsection{Central pixel}
No sampling error can be computed for the single central
pixel. In this case we apply the Poisson error, noting that the
central pixel value has been scaled up from a very short (0.26
seconds) snapshot exposure in order to avoid saturation. The error
on the central pixel is typically of the same order as, or a little
larger than, the sampling error deduced for the innermost annulus.

\subsection{Minimization}
The model quasar was convolved with the PSF and, after masking of
diffraction spikes and nearby companion objects, compared with the HST
data. The $\chi^2$ minimum was found for each object within the
6-dimensional parameter space using the downhill simplex method
\citep{numrec}. Six different starting points (spanning the entire
parameter space) were used for each object to ensure that the global
minimum had been located.  
This model was checked using three starting simplexes
at $\pm$2, 5, and 10\% of each best-fitting parameter value. Finally,
the model was double-checked by performing a grid search
around the best-fit model.

\subsection{Photometry}
Photometric calibration was performed using the HST headers {\sc
  PHOTFLAM} and {\sc PHOTZPT}, in order to convert counts into
physical 
units of spectroscopic flux density (erg s$^{-1}$ cm$^{-2}$
\AA$^{-1}$). In each case, the rest frame filter band is calculated and
compared to standard Johnson $V$-band. 
We assume a simple flat spectrum across the filter bandpass.  The
internal uncertainty in photometry due to the accuracy of the
calibration reference files and the stability of the instrument is
1-2\%.  The conversion from the HST to the Johnson photometric system
also has an uncertainty of a few percent.


\section{Results}
As described in the previous section, we used two separate modelling
strategies in order to determine the morphology of the hosts and the
relative contributions of the nuclear and galaxy components. In the
first case we fitted a pure de Vaucouleurs ($r^{1/4}$-law) elliptical
galaxy and then a pure (exponential) Freeman disc to the host and used
the difference in the $\chi^{2}$ values for the two models to decide
which model gave the best overall fit. Unless otherwise stated in the
notes, all objects were modelled out to a radius of 4 arcsec.
Table 3 lists the results of this strategy.

\begin{table*}
  \caption{Results of model fitting with both de Vaucouleurs spheroid
    and Freeman disc models. Columns are as follows: object name; best
    fitting host-galaxy morphology (disc or elliptical);
    reduced-$\chi^2$ value for the best fit model; $\Delta \chi^2$
    between the chosen and alternative-morphology model; half-light
    radius, $R_{1/2}$, of best fitting galaxy model in kpc; surface
    brightness of the host at the half-light radius, $\mu_{1/2}$, in
    units of $V$ mag.arcsec$^{-2}$; integrated absolute magnitudes of
    the nucleus and the host galaxy, converted from the appropriate
    filter band (F814W/F791W) to Johnson $V$-band; the ratio of
    integrated nuclear and host galaxy luminosities; position angle of
    the host (in degrees, anti-clockwise from vertical in the images);
    the axial ratio of the host.}  
  \begin{tabular}{clcrrcccrrc}
    \hline
    \hline
    IAU name&Morphology&$\chi^{2}_{red}$&$\Delta\chi^{2}$& 
    $R_{1/2}$&$\mu_{1/2}$&$M_{V}^{nuc}$&$M_{V}^{host}$&$L_{N}/L_{H}$&PA&$a/b$\\
    &(best fit)& & & (kpc)& &$$& & &($^{\circ}$)& \\
    \hline
    \multicolumn{11}{c}{Radio-Quiet Quasars}\\
    0624$+$691 &Elliptical&$1.488$&$486.2$&$9.7 \pm0.7$&$22.0\pm0.15$&$-27.18$&$-24.01$&18.45&140&1.2\\
    1001$+$291 &Elliptical&$1.318$&$326.3$&$15.4\pm0.6$&$23.1\pm0.10$&$-25.62$&$-23.47$&7.27 & 58&1.7\\
    1230$+$097 &Elliptical&$1.219$&$107.7$&$5.8 \pm0.2$&$21.7\pm0.13$&$-25.24$&$-23.24$&6.26 &  1&1.3\\
    1237$-$040 &Disc      &$1.323$&$334.5$&$6.7 \pm0.1$&$22.0\pm0.05$&$-23.98$&$-22.63$&3.46 & 14&1.1\\
    1252$+$020 &Elliptical&$1.228$&$102.0$&$3.9 \pm0.5$&$22.1\pm0.30$&$-25.26$&$-22.04$&19.47&150&1.1\\
    1254$+$021 &Elliptical&$1.356$&$646.3$&$14.2\pm0.3$&$23.2\pm0.05$&$-23.91$&$-24.00$&0.92 & 30&1.1\\
    1258$-$015 &Elliptical&$1.352$&$ 14.4$&$1.5\pm0.2$&$19.8 \pm0.25$&$-23.77$&$-22.33$&3.77 &140&1.1\\
    1313$-$014 &Disc      &$1.254$&$351.0$&$5.6 \pm0.1$&$21.7\pm0.06$&$-23.74$&$-22.68$&2.65 &174&1.1\\
    1357$-$024 &Disc      &$1.257$&$457.1$&$5.8 \pm0.1$&$21.9\pm0.05$&$-23.66$&$-22.47$&2.99 &160&1.2\\
    1821$+$643 &Elliptical&$1.828$&$570.4$&$18.9\pm0.2$&$22.9\pm0.05$&$-27.14$&$-24.33$&13.35&113&1.3\\
    \hline
    \multicolumn{11}{c}{Radio-Loud Quasars}\\
    0031$-$707 &Elliptical&$1.268$&$802.8$&$11.0\pm0.4$&$23.1\pm0.08$&$-23.84$&$-23.21$& 1.80& 70&1.2\\
    0110$+$297 &Elliptical&$1.327$&$561.1$&$12.3\pm0.8$&$23.6\pm0.12$&$-23.93$&$-22.99$& 2.39& 31&1.2\\
    0812$+$020 &Elliptical&$1.509$&$904.7$&$17.4\pm0.3$&$23.6\pm0.05$&$-24.80$&$-23.81$& 2.49&155&1.2\\
    1058$+$110 &Elliptical&$1.390$&$135.5$&$13.1\pm1.1$&$24.1\pm0.14$&$-23.58$&$-22.69$& 2.28&159&1.3\\
    1150$+$497 &Elliptical&$1.360$&$340.6$&$8.3 \pm0.3$&$22.1\pm0.11$&$-24.09$&$-23.28$& 2.11&174&1.5\\
    1208$+$322 &Elliptical&$1.096$&$ 49.1$&$6.5 \pm0.1$&$22.2\pm0.05$&$-25.01$&$-22.50$&10.06&  7&1.9\\
    1233$-$240 &Elliptical&$1.358$&$ 47.7$&$3.1 \pm0.1$&$20.6\pm0.05$&$-24.78$&$-22.98$& 5.23& 58&1.1\\
    \hline
    \hline
  \end{tabular}
\end{table*}

\begin{figure*}
  \label{fig-prof}
  \centering
  \includegraphics[width=168mm]{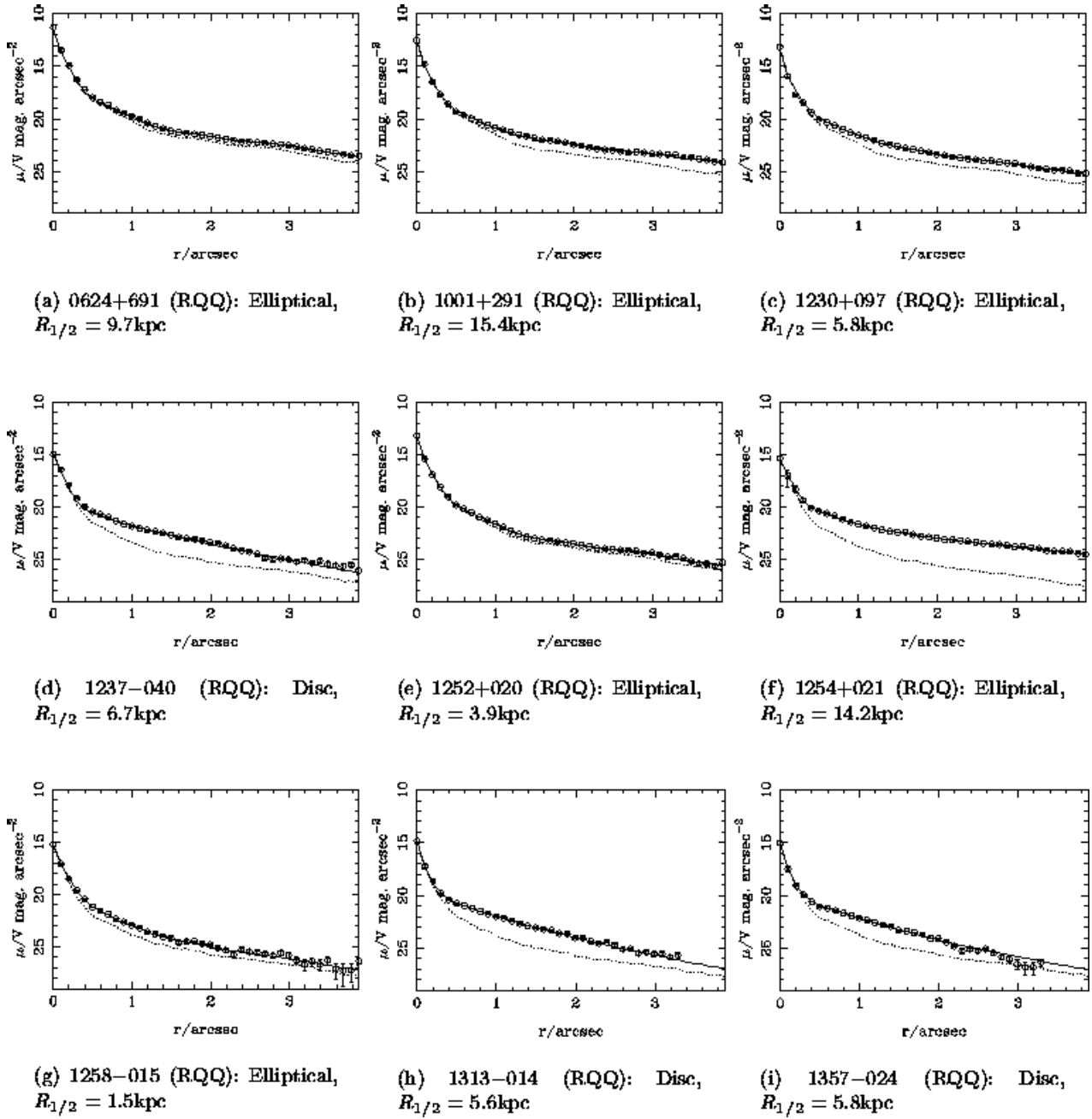}
  \caption{The radial profiles of the best-fitting bulge or disc models
    for the 17 quasars in our sample. Each plot shows the azimuthally
    averaged image data (open circles with $1\sigma$ error bars), the
    azimuthally averaged best-fit model after convolution with the PSF
    (solid line) and the azimuthally averaged best-fit unresolved nuclear
    component after convolution with the PSF (dotted line). The form of
    the fit (disc or elliptical) and the scale length ($R_{1/2}$) of the
    model galaxy are also given beneath each panel.}
\end{figure*}

\addtocounter{figure}{-1}
\begin{figure*}
  \centering
  \includegraphics[width=168mm]{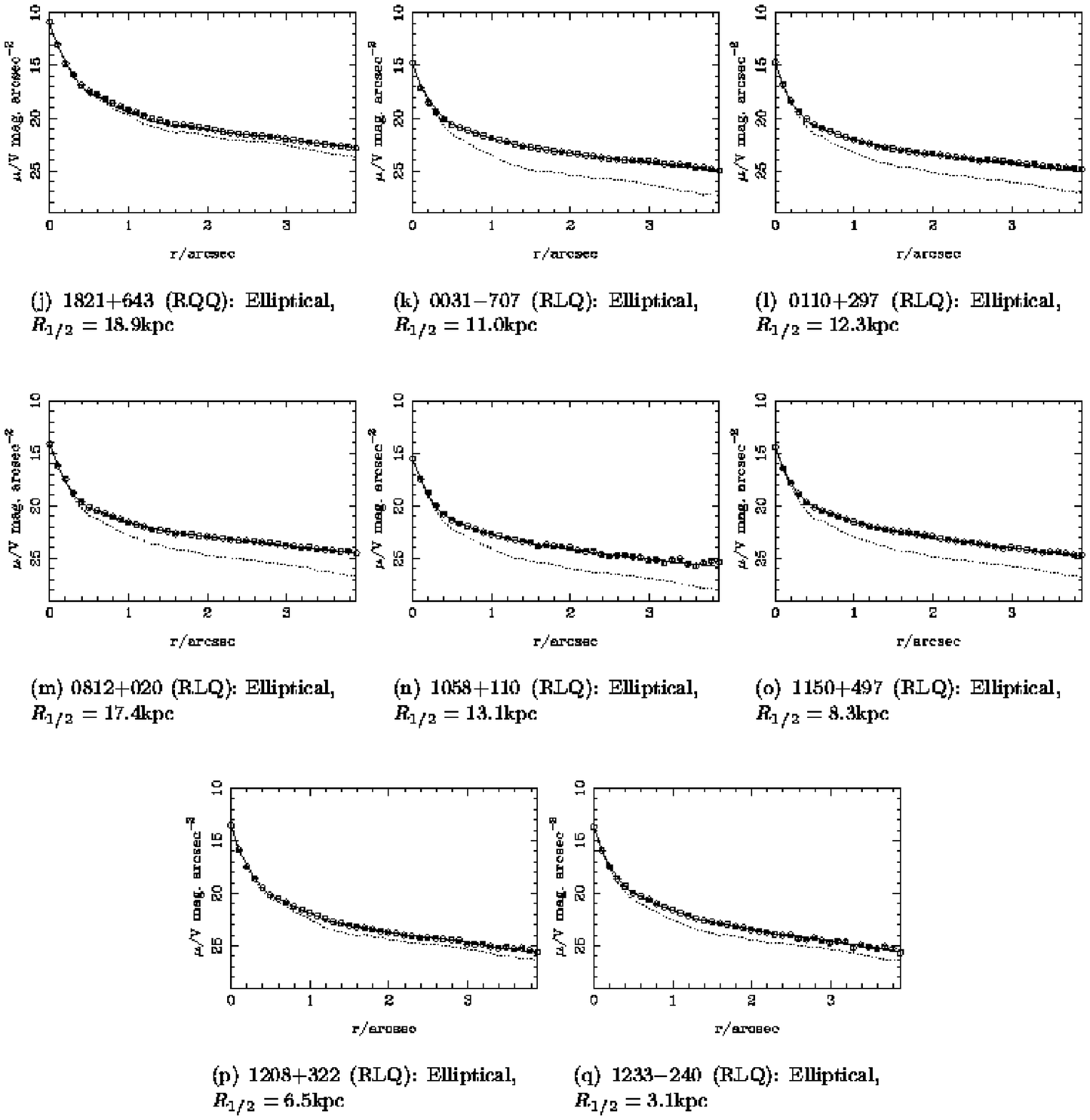}
  \caption{\em - continued}
\end{figure*}

In the second case, we carried out modelling using a variable-$\beta$
fit, in which the $\beta$ parameter of equation 1 is allowed to vary
freely. This allows for a more
general morphology than the strictly disc or bulge technique. Table 4
shows the results of this variable-$\beta$ fitting which, for the most part, 
rather impressively reinforces the results of the fixed-$\beta$ models. 
However there are a few objects in which the variable-$\beta$
technique returned a hybrid value and these are noted in the entry for
the relevant object in the Appendix.

Greyscale images of the individual objects are also presented in the
Appendix. For each quasar we show the final reduced $I$-band
(F814W/F791W) HST image (top left), the best-fitting model quasar
(nuclear point source plus either pure bulge or pure disc host galaxy)
to the quasar image (top right), the model host
galaxy only (bottom left) and the model-subtracted residual image
(bottom right). Radial profiles for the best-fit bulge and disc models
are presented in Fig.2.

\begin{figure*}
  \label{fig-cont}
  \centering
  \includegraphics[width=75mm, angle=270]{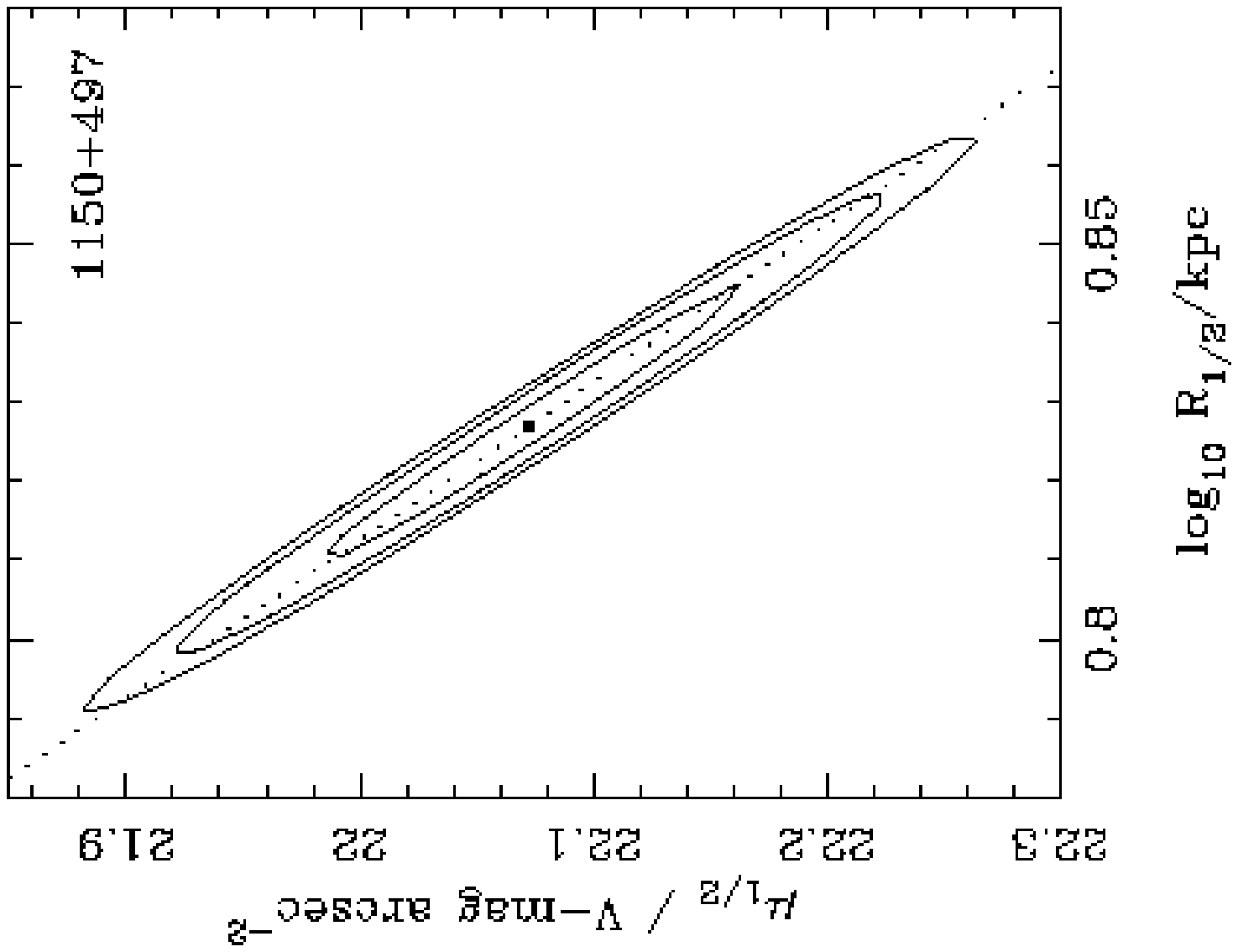}
  \includegraphics[width=75mm, angle=270]{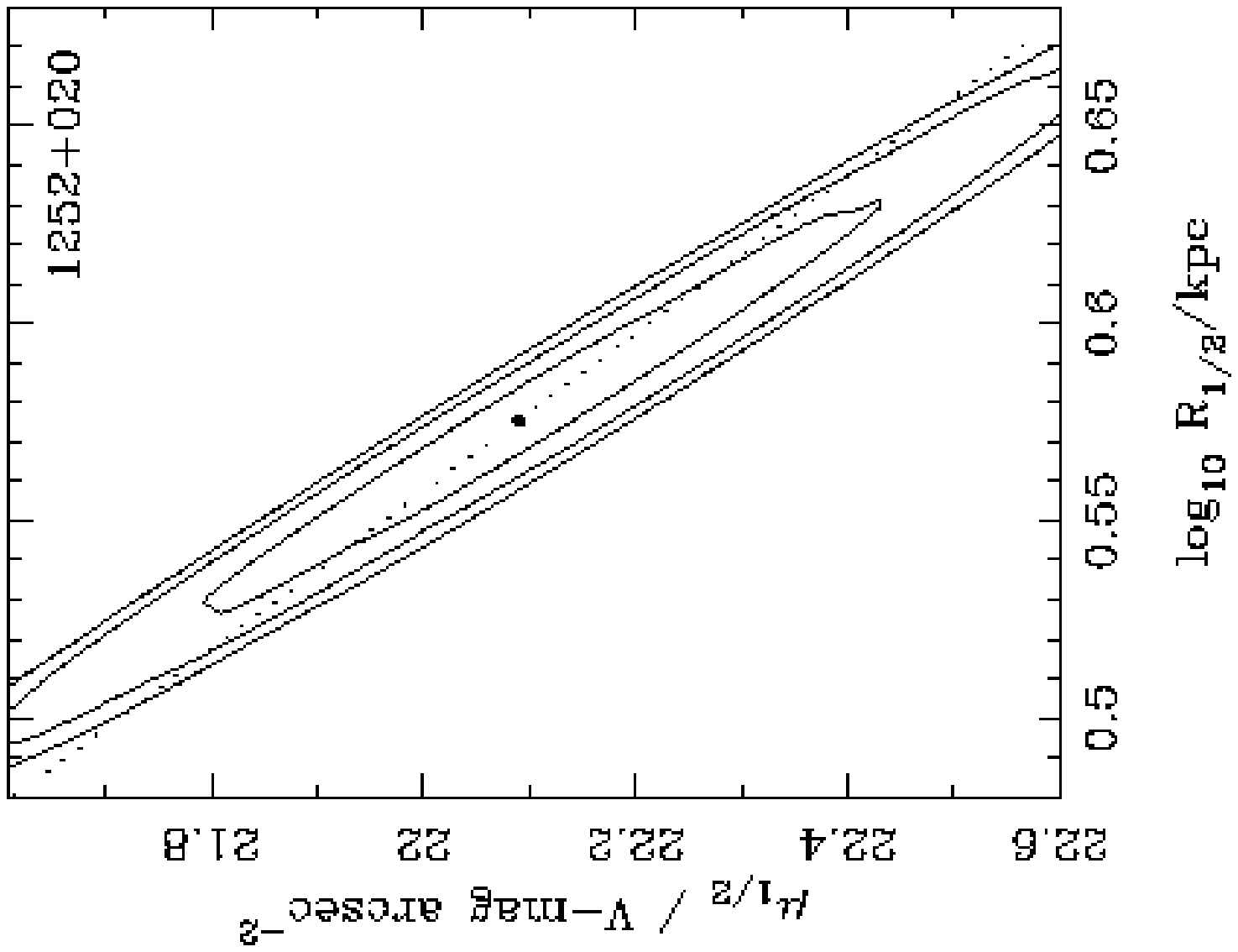}
  \caption{$\chi^2$ contours (at $1-3\sigma$ levels) in the $\mu - R$
    plane for 1150$+$497 (left), and 1252$+$020 (right). 
    In each case, the best-fit
    solution is marked by a dot. The figure demonstrates the
    degeneracy that remains between galaxy size and surface brightness,
    even when any confusion between host and nuclear light has been
    eliminated. 1150$+$497 is a typical example of the objects in
    this study and the contours describe a slope close to 5 (dashed
    line), as expected if the host galaxy's luminosity has been
    correctly constrained. Of all the objects in our sample, 1252$+$020
    has the least robust host luminosity constraint, the contours lying along a
    somewhat steeper slope of $\simeq5.7$. One other object, 1258$-$015,
    exhibits slightly steep contours, having a slope of $\simeq
    5.6$. In these latter 2 cases there is clearly still some degeneracy
    between host and nuclear flux.}
\end{figure*}

We can gain further insight into how successful we have been
in disentangling the host galaxy from the nucleus through
investigation of the $\chi^2$ contours in the $\mu_{1/2}-R_{1/2}$ 
plane (e.g. Fig.3).
For any quasar in which we have successfully characterised the
host luminosity (i.e.  eliminated the degeneracy between host and
nuclear contributions), these contours should lie along a slope of 5.0
(see e.g.~\citealt*{abraham+92, malkan84}), and allow us to assess how
well constrained these two parameters are.

\begin{table}
  \caption{Outcome of variable-$\beta$ modelling. Columns are as
    follows: object name; best-fit morphology from pure bulge \& disc
    models (Table 3); best-fit $\beta$ value with no assumed morphology;
    the value of reduced-$\chi^2$ produced by this best-fit
    model; improvement in fit, $\Delta \chi^2$
    obtained by using the variable-$\beta$ technique compared to the
    best-fit fixed morphology model.}
  \centering
  \begin{tabular}{clccr}
    \hline
    \hline
    IAU name&Morph.&$\beta$&$\chi^{2}_{red}$&$\Delta\chi^{2}$\\
    \hline
    \multicolumn{5}{c}{Radio-Quiet Quasars}\\
    0624$+$691 &Elliptical&0.20&1.485&  42.2\\
    1001$+$291 &Elliptical&0.26&1.318&   0.1\\
    1230$+$097 &Elliptical&0.37&1.216&  12.0\\
    1237$-$040 &Disc      &0.96&1.321&   0.6\\
    1252$+$020 &Elliptical&0.22&1.227&   0.7\\
    1254$+$021 &Elliptical&0.24&1.356&   2.0\\
    1258$-$015 &Elliptical&0.26&1.351&   2.0\\
    1313$-$014 &Disc      &0.97&1.254&   0.2\\
    1357$-$024 &Disc      &1.32&1.246&  22.8\\
    1821$+$643 &Elliptical&0.22&1.771& 485.4\\
    \hline
    \multicolumn{5}{c}{Radio-Loud Quasars}\\
    0031$-$707 &Elliptical&0.26&1.266 &11.9\\
    0110$+$297 &Elliptical&0.22&1.323 &15.5\\
    0812$+$020 &Elliptical&0.23&1.503 & 9.5\\
    1058$+$110 &Elliptical&0.33&1.388 & 6.9\\
    1150$+$497 &Elliptical&0.36&1.356 &19.4\\
    1208$+$322 &Elliptical&0.32&1.091 &19.5\\
    1233$-$240 &Intermediate&0.56&1.361&57.2\\
    \hline      
    \hline      
  \end{tabular}
\end{table} 
\begin{figure}
  \centering
  \includegraphics[width=80mm]{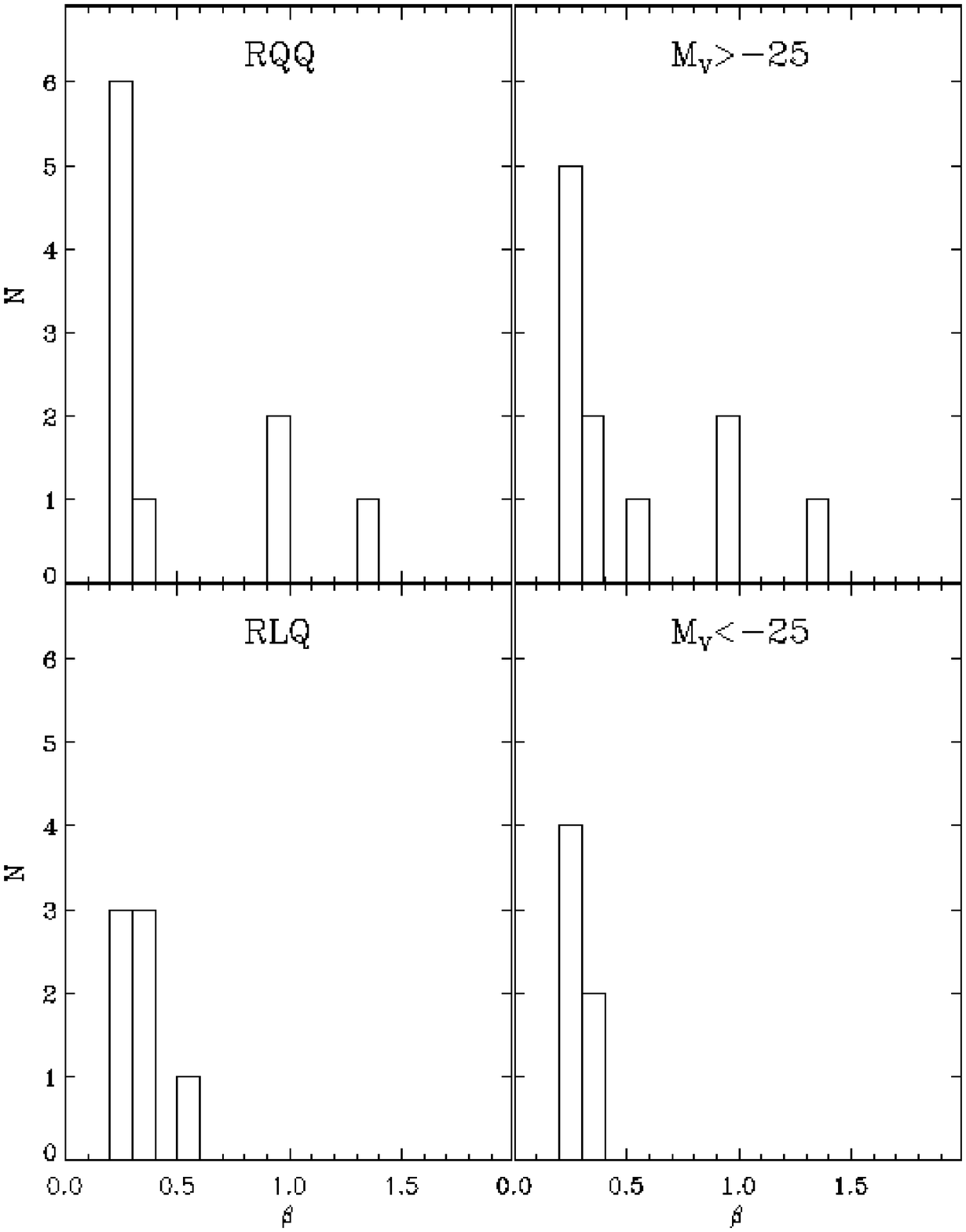}
  \caption{Histograms of the best-fit $\beta$ values from our
    variable-$\beta$ models. $\beta=0.25$ is equivalent to an $r^{1/4}$ de
    Vaucouleurs law, $\beta=1$ is an exponential Freeman disc. In the
    panel to the left we divide the sample in terms of radio
    luminosity. On the right, the sample is divided according to optical
    absolute magnitude. Clearly all optically-luminous, and all radio-loud
    objects lie in bulge-dominated hosts, confirming and extending the trends
    deduced by Dunlop et al. (2003).}
\end{figure}


\section{Discussion}
\label{sec-disc}
The quasars imaged in this study span
almost two orders of magnitude in optical luminosity but only a
narrow range of redshifts. They therefore allow us to investigate the
relationship between galaxies and their central black-holes, and the
relative roles of black-hole mass and fuelling efficiency in
determining quasar luminosities.

We have successfully recovered a host galaxy for each one of the
17-strong sample. In general, the host size and central surface
brightness have been constrained to within a few kiloparsecs, and half a
magnitude, respectively, as is illustrated by the typical joint confidence 
region illustrated in the left-hand panel of Fig.3.  
However, there are two objects, the RQQ's 1252$+$020, and 1258$-$015
for which the fit yields poor stability for the host properties, and
the resulting much larger confidence region for 1252$+$020 is
shown in the right-hand panel of Fig.3.
Overall host and nuclear fluxes
are typically constrained to better than 0.1mag by the modelling
software, with a similar uncertainty due to the conversion from ST mags to
standard $V$-band.

\subsection{Host galaxy morphologies}
With regard to basic host galaxy morphology, the results of this study are 
quite clear cut, and confirm and extend the findings of
\citet{dunlop+03}. 
For every quasar host the modelling software yielded a clear 
decision in favour of either a disc-dominated or bulge-dominated
host. Moreover, in virtually every case this preference was confirmed
by the variable-$\beta$ model, which returned a value of $\beta$ very
close to either $0.25$ (elliptical) or $1$ (disc). 

At this point it is important to clarify what we mean by
``bulge-dominated'' or ``disc-dominated'' galaxies.
Our modelling software finds the best overall fit to the light from the
quasar and its host galaxy. This is dominated by contributions from the
point-like nucleus itself, and from the smooth, high SNR host region far
from the nucleus.  Our HST images are of sufficient depth that we expect
to be able to detect large features at least as dim as $V=27$
mag.arcsec$^{-2}$. 
To place this in some context, this is a sufficient depth that the
prominent tidal arm in Mrk1014, could be easily be detected if it were
placed at a $z\sim1$.
\footnote{Markarian 1014 is a well-known disturbed active galaxy at
  $z=0.163$, with a prominent tidal arm that is easily detected at a
  surface brightness of $\simeq24$ V mag. arcsec$^{-2}$
  (see images and profiles in McLure et al. 1999).}
However, close to the nucleus, this sensitivity is impaired by our lack of
knowledge of the form of the PSF, and here a small-scale, relatively
bright feature might go unnoticed by the model. Such features can be
exposed in our residual images, which show the best fit model subtracted
from the data. Thus when we claim to detect bulge or disc-dominated hosts,
we mean just that: on large scales of a few kpc, the host light is
dominated by smooth emission that follows a spheroidal ($r^{1/4}$-law) or
a disc-like (exponential) profile.
There obviously remains the possibility of a small, centrally
concentrated bulge component in the disc-dominated hosts, and in an
effort to address this, we attempted a 2-component modelling
technique. This is simply an extension of the modelling technique
described above, with individual disc and bulge components each
described in full by 4 parameters, giving a 9-dimensional model
overall. The results are presented in table 5:
A significant spheroidal component is found in two of the three
disc-dominated hosts; 1237$-$040 and 1313$-$014.

\begin{table*}
  \caption{Results of the 2-component model fitting to the
  disc-dominated hosts. The improvement in the fit over the simple
  disc case is given in column 3. 
  The nuclear luminosity, and the scale-length and luminosity of each
  host galaxy component are given in columns 4-8, with the bulge: disc
  ratio in the final column.} 
  \begin{tabular}{ccrrllrrc}
    \hline
    \hline
    Source&Morph.&$\Delta\chi^{2}$&$M_{V}(\mathrm{Nuc})$&$r_{B}$&
    $M_{V}(\mathrm{Bulge})$&$r_{D}$&$M_{V}(\mathrm{Disc})$&Bulge/Disc\\
    & & & &kpc& &kpc& & \\
    \hline
    1237$-$040&D/B& 10.8&$-23.96$& 3.8&$-21.18$& 6.5&$-22.46$&0.31\\
    1313$-$014&D/B&  4.7&$-23.72$& 3.3&$-20.28$& 5.6&$-22.62$&0.12\\
    1357$-$024&D  &  0.3&$-23.65$& 0.4&$-18.38$& 6.3&$-22.47$&0.023\\
    \hline
    \hline
  \end{tabular}
\end{table*}

As illustrated in Fig.4, the three quasars in the current sample for which
we find disc-dominated hosts are, i) radio-quiet, and ii) in the
low-luminosity sub-sample with $M_V > -25$. In fact, reference to
Table 3 reveals that the three disc-dominated hosts house nuclei with
$M_V > -24$. Furthermore, the brightest of the three quasars
(1237$-$040), is found to have a significant spheroid. 
This result therefore meshes
well with the luminosity-dependence of host-galaxy morphology
illustrated in Fig.10 of~\citet{dunlop+03}; disc-dominated host
galaxies are not found for nuclei with $M_V < -24$. 
In one other case (1313$-$014) we detect a spheroidal component at low
luminosity ($M_{V}=-20.28$), suggesting an accretion rate around
$L/L_{Edd}\simeq3$, assuming the local $M_{BH}-M_{\mathrm Sph}$ ratio
\cite{mcluredunlop02}.
In the final disc-dominated object (1357$-$024) we detect no
significant bulge component. However, for a roughly Eddington-limited
accretion rate, we would anticipate a bulge luminosity of around 
$M_{V}=-20.8$. Such a bulge is expected to have a scalelength of
$\simeq3$kpc, and could be difficult to distinguish from the
unresolved nuclear light.

As discussed by \citet{dunlop+03}, this result can now be viewed as
a natural consequence of the now well-established proportionality of
black-hole and spheroid mass. 

\subsection{Host galaxy scale lengths and luminosities}
Table 3 lists the scale lengths and luminosities for the best-fit
fixed-morphology models. Once again, the results are broadly
consistent with those of \citet{mclure+99} and \citet{dunlop+03};
the hosts are generally large, luminous galaxies. 

Three of the five smallest galaxies are the disc-dominated hosts.
There is a tendency for the hosts of the RLQs to be slightly
larger than those of the RQQs, but this is not statistically significant.

\[\langle R_{1/2}\rangle_{(RLQ)}=10.2 \pm 1.8  \mathrm{kpc} \]
\[\langle R_{1/2}\rangle_{(RQQ)}=8.7 \pm 1.8  \mathrm{kpc} \]

On average the more luminous quasars also have slightly larger hosts,
but again the mean values for the two subsamples are in agreement 
given the statistical uncertainty.

\[\langle R_{1/2}\rangle _{(M_{N}<-25)}=10.0 \pm 2.4 \mathrm{kpc}\]
\[\langle R_{1/2}\rangle _{(M_{N}>-25)}=9.0 \pm 1.5 \mathrm{kpc}\]

We find that that all the hosts are more luminous than $L^{\star}$
($M_{V}^{\star}=-21.0$;~\citealt*{EEP88}). There is no statistically 
significant difference between the average 
values for each subsample, but these basic statistics should not 
obscure the fact that the two quasars in the sample with $M_V(Nuc) < -27$
are also the only two objects for which we find $M_V(Host) < -24$.

\[\langle M_{V}(Host)\rangle _{(RLQ)}=-23.06 \pm 0.16 \]
\[\langle M_{V}(Host)\rangle _{(RQQ)}=-23.12 \pm 0.25 \]
\[\langle M_{V}(Host)\rangle _{(M_{N}<-25)}=-23.27 \pm 0.36 \]
\[\langle M_{V}(Host)\rangle _{(M_{N}>-25)}=-23.01 \pm 0.16 \]

\subsection{Kormendy relation}
The Kormendy relation is the photometric projection of the fundamental
plane exhibited by elliptical galaxies.
The host galaxies of the quasars in our sample follow
a Kormendy relation of the form
\begin{equation}
\mu_{1/2}=(19.2\pm0.6)+(3.33\pm0.7)log_{10}R_{1/2}
\end{equation}
shown in Fig.5 (where we have plotted and fitted only those with
bulge-dominated hosts).  A galaxy with a well-constrained luminosity
but unknown scale length will lie along a locus with a slope of 5,
illustrated by the error ellipse in the top right corner of this figure
(c.f. Fig.3).  The slope of 3.33 is in excellent agreement with that
determined recently for 9000 early-type galaxies ($3.33\pm0.09$) drawn
from the SDSS by \citet{bernardi+03}
and is sufficiently different to a slope of 5 to convince us that the
surface brightnesses and scale lengths of the hosts have been well
constrained. 

\begin{figure}
  \centering
  \includegraphics[width=70mm]{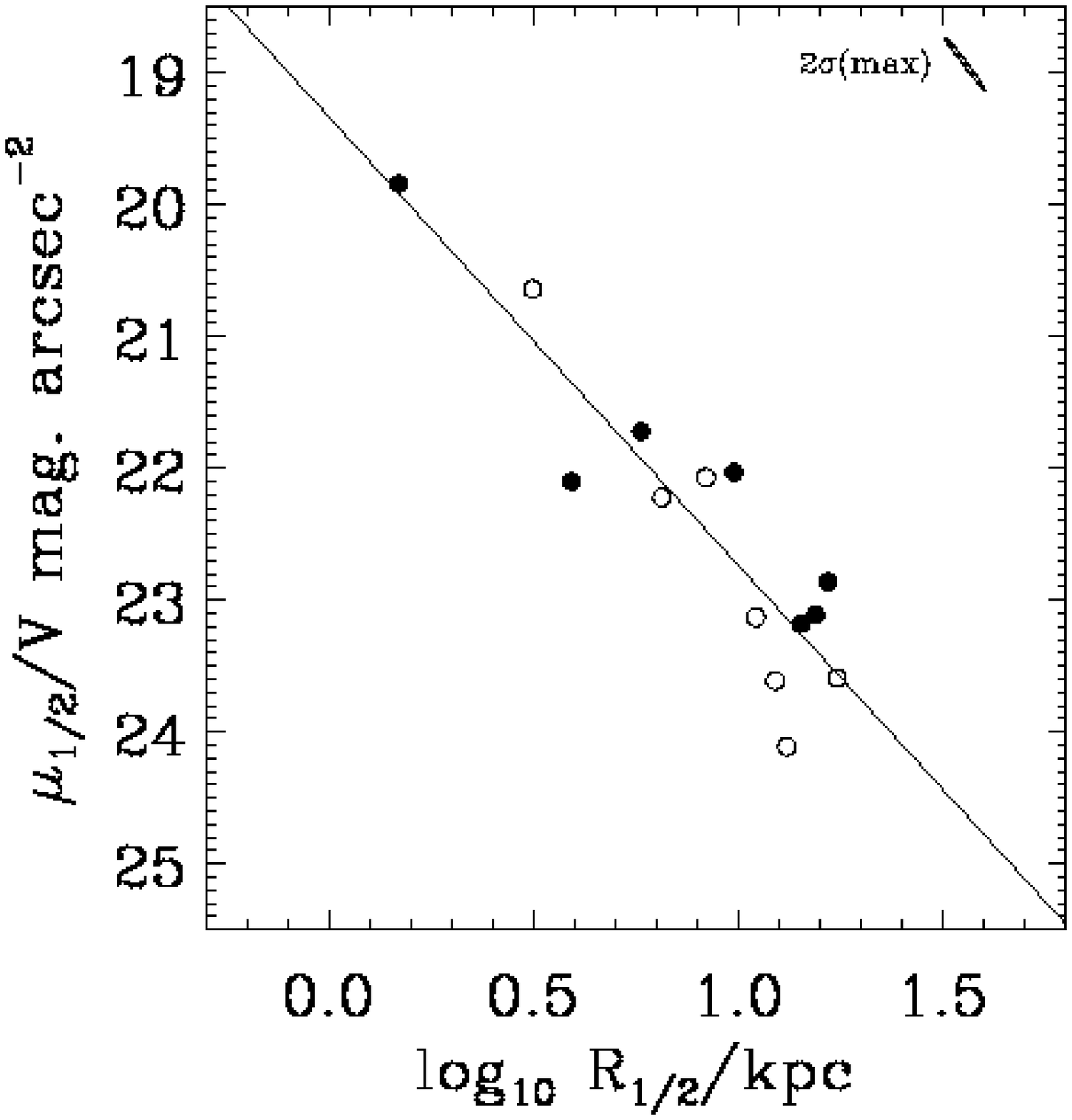}
  \caption{The scale length ($R_{1/2}$) vs surface brightness
    ($\mu_{1/2}$) projection of the fundamental plane. Filled circles
    represent RQQ hosts, open circles RLQ hosts (bulge-dominated hosts
    only). The solid line shows the
    best fit Kormendy relation to the sample and has the form
    $\mu_{1/2}=(19.2\pm0.6)+(3.33\pm0.7)log_{10}R_{1/2}$.
    The narrow ellipse in the top
    right corner of the plot shows the $2\sigma$ error contours for
    1258$-$015; its slope of 5 is due to the remaining degeneracy between size
    and surface brightness when host luminosity has been well constrained.} 
\end{figure}

\subsection{The role of galaxy mergers and interactions}
Interactions and merging events between galaxies have long been
suggested as the triggering events for the activation of quasars,
especially at low redshifts where some mechanism is required to
initiate fuelling of the black-holes in otherwise stable, gas-depleted
ellipticals. Indeed, most host galaxy studies to date have found that
indications of disturbance such as tidal tails, multiple nuclei and
close companions are present in around 50\% of quasar hosts
(e.g. \citealt{smith+86,hutchingsneff92,bahcall+97}).
However~\citet{dunlop+03} point out that this is also true of inactive
massive ellipticals, so that it is not clear whether mergers are
genuinely a defining feature of quasar hosts or merely the legacy of
their parent population.  Certainly many quasar hosts appear to be
entirely undisturbed, so clearly a large-scale disruption of the host
is not always necessary to trigger fuelling of the central engine (or
at least the timescales for relaxation after the merger event and of
fuel reaching the central engine may sometimes be vastly different).

The residual images of the quasars in the current study (see
Appendix), provide a means of identifying signs of galaxy interactions
which might not be obvious in the raw HST images. They are produced by
subtracting the best-fitting axially-symmetric 
quasar model from the HST image. Since the model
only attempts to fit the smooth underlying distribution of galaxy
light, any additional structures (spiral arms, bars, tidal tails,
double nuclei etc) will be made more obvious in the residual image.

In our sample we find unambiguous evidence for an ongoing galaxy
interaction in only one object, the RQQ 1237$-$040. Several other
objects are candidates for some form of disturbance having taken place
(for instance the RQQ 1001$+$291 with its prominent spiral arms and
large-scale de Vaucouleurs profile), or have other objects within a
few arcsec on the sky which might conceivably be interacting if they
lie at the same redshift. In fact the majority of our quasars appear
to have some companions nearby on the sky, which is at least
suggestive of a cluster environment.

However we find no correlation between the luminosity of the quasar
and the presence of any morphological disturbance in the host. In our
small sample, at least, the most luminous quasars seem no more likely
to be interacting systems than their less luminous counterparts.

\subsection{Black-hole masses}
Reliable black-hole masses are available for at least 37 nearby
galaxies \citep{kormgeb01}, highlighting the correlations which exist
between spheroid luminosity and black-hole mass
(e.g.~\citealt{magorrian+98}), and between spheroid
velocity-dispersion and black-hole mass \citep{gebhardt+00,
ferraresemerritt00}. As discussed by \citet{mcluredunlop02}, both these 
relations for inactive ellipticals, and the results of H$\beta$-derived virial
black-hole estimates in active objects, are consistent with a direct 
proportionality of black-hole and spheroid mass of the form
$M_{BH}=0.0012 M_{\mathrm Sph}$, with a typical scatter of 0.3 dex.

We used the luminosities from our best-fit bulge-dominated galaxy
models to estimate host-galaxy masses, given an estimate of the
mass-to light ratio of an early-type galaxy \citep*{jorgensen+96}; $
(M/L)_{R-band} \propto L^{1.31}$.
We then used the black-hole:spheroid mass relation above to estimate
black-hole masses for the quasars, independent of their observed 
nuclear output. 
The results of this calculation are given in column 3 of Table 6.
Once again, this has only been performed for the bulge-dominated
hosts. A similar calculation can be performed for the spheroidal
components of the disc-dominated hosts (see end of section 7.1). 
The median black-hole masses for the high and low-luminosity
subsamples are: 
\[\langle M_{BH})_{(M_{N}<-25\rangle }=(1.1\pm1.0)\times10^{9}\]
\[\langle M_{BH})_{(M_{N}>-25\rangle }=(0.5\pm0.3)\times10^{9}\]
We find that all of the host galaxies are sufficiently massive
($M_{\mathrm Sph}>10^{11}M_{\odot}$) to contain a black-hole in excess of
$10^{8}M_{\odot}$, but the difference in mass between the black-holes
in optically powerful and optically weak quasars is not large enough
to account for the factor $\sim 10$ increase in luminosity.
Unfortunately, no direct measures of black-hole mass are available for
any of the objects in our sample. A comparison with, for example, a
Virially estimated black-hole mass would provide an extremely valuable
test.

\subsection{Fuelling efficiencies}
We can now calculate the predicted luminosity of each bulge-dominated
object if the black-hole were to radiate at its Eddington limit
($L_{Edd}^{Bol} = 1.26 \times 10^{31} \frac{M_{BH}}{M_{\odot}}$ Watts)
and compare this with the actual luminosity of
the quasar nucleus obtained from our model fitting. The results of this
procedure are listed in columns 4 and 5 of Table 6, and plotted in
Fig.6; there is clearly no correlation between black-hole mass and
fuelling efficiency. However, one can immediately see a clearer
distinction between our luminous and dim subsamples than was apparent
simply from their estimated black-hole masses:

\[\langle\frac{L_{N}}{L_{Edd}}\rangle_{(M_{N}<-25)} = 0.78\pm0.2\]
\[\langle\frac{L_{N}}{L_{Edd}}\rangle_{(M_{N}>-25)} = 0.17\pm0.03\]
If we exclude the relatively poorly constrained luminous RQQ
1252$+$020 (which appears, from our modelling, to exceed the Eddington
limit), the median Eddington ratio 
for the luminous subsample drops to 0.47, but this is clearly
still significantly higher than for the low-luminosity sample.

\begin{table}
  \begin{tabular}{lcrrrr}
    \hline
    \hline
    Object&Morph.&$M_{\mathrm Sph}$&$M_{BH}$&$M_{V}^{Edd}$&$\frac{L_{N}}{L_{Edd}}$\\ 
    & &($10^{11}M_{\odot}$)&($10^{9}M_{\odot}$)& & \\ 
    \hline
    \multicolumn{6}{c}{Radio-Quiet Quasars}\\
    0624$+$691&E  &13.90&1.67&$-27.28$&0.91\\
    1001$+$291&E  & 7.27&0.87&$-26.57$&0.42\\
    1230$+$097&E  & 5.51&0.66&$-26.27$&0.39\\
    1252$+$020&E  & 1.29&0.16&$-24.70$&1.67\\
    1254$+$021&E  &13.70&1.65&$-27.27$&0.05\\
    1258$-$015&E  & 1.84&0.22&$-25.08$&0.30\\
    1821$+$643&E  &20.50&2.46&$-27.70$&0.60\\
    \hline 
    \multicolumn{6}{c}{Radio-Loud Quasars}\\
    0031$-$707&E  & 5.31&0.64&$-26.23$&0.11\\
    0110$+$297&E  & 4.07&0.48&$-25.94$&0.16\\
    0812$+$020&E  &11.00&1.31&$-27.02$&0.13\\
    1058$+$110&E  & 2.84&0.34&$-25.55$&0.16\\
    1150$+$497&E  & 5.78&0.69&$-26.32$&0.13\\
    1208$+$322&E  & 2.26&0.27&$-25.30$&0.76\\
    1233$-$240&E  & 4.02&0.48&$-25.93$&0.35\\
    \hline
    \hline
  \end{tabular}
  \caption{Galaxy spheroid and black-hole mass estimates for 
    each of the bulge-dominated quasars in our sample. The
    table also lists the theoretical Eddington luminosity, $M_{V}^{Edd}$,
    of each black-hole, and the efficiency at which the black-hole is
    accreting expressed as the ratio of the luminosity ascribed by our
    model to the nuclear point source to the Eddington luminosity
    predicted by the model of the host galaxy ($L_{N}/L_{Edd}$). Note
    that the RQQ 1252$+$020 appears to be accreting at a super-Eddington
    rate. This object has the least robust model fit of the entire sample,
    and it is likely that the nuclear flux has been
    overestimated. Results for the spheroidal components of the three
    disc-dominated objects are presented where detected.
    Both detections (1237$-$040 and 1313$-$014) yield slightly
    super-Eddington accretion rates, whilst the nuclear luminosity of
    1357$-$024 can be explained in terms of Eddington-limited
    accretion in an undetected, compact central bulge at
    $M_{V}\simeq-20.8$.}
\end{table}

\begin{figure}
  \centering
  \includegraphics[width=70mm]{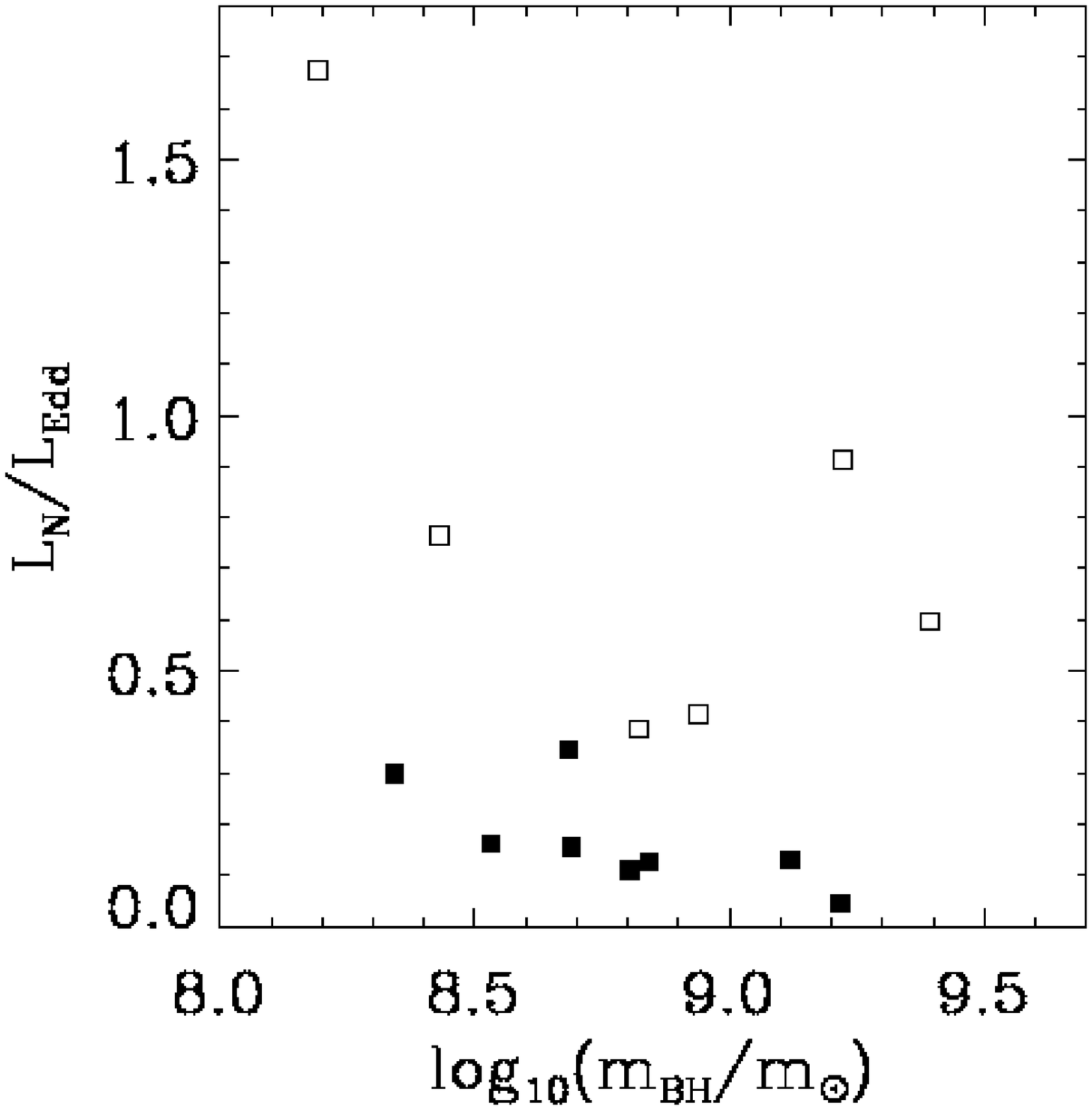}
  \caption{Quasar efficiency as a function of the Eddington luminosity
    versus black-hole mass, as determined from the spheroid
    luminosity. Only bulge-dominated objects are plotted, and they have 
    been divided into
    optically luminous (open squares) and optically dim (solid
    squares) subsamples at $M_{V}(Nuc)=-25$, as discussed in the
    text. Our model fit to the RQQ 1252$+$020 implies that the object
    has a small host with an extremely luminous nucleus, yielding a
    super-Eddington luminosity. However, the quality of the model fit
    is poor, and this value is quite likely erroneous. Overall there
    is no obvious tendency for fuelling efficiency to vary as a
    function of black-hole mass, and the increase in luminosity across
    the sample reflects an increase both in black-hole mass, and
    accretion efficiency.}
\end{figure}

Thus, within our $z = 0.4$ sample, increasing quasar luminosity appears,
on average, to reflect a mix of both larger black-hole mass, and increased 
fuelling efficiency. Not surprisingly, the only two quasars in the present 
sample with $M_V < -27$ have the two most massive black-holes.

A number of other features of the results summarised in Table 6 are worthy
of comment. First, while inferred fuelling efficiencies range over an order 
of magnitude, we find no evidence for super-Eddington accretion. If
one excludes the poorly constrained RQQ 1252$+$020, the most efficient
emitter is 0624$+$691, with $L_{N}/L_{Edd} \simeq 1$.
Second, the most massive central black-hole found in our sample has a
mass of $3 \times 10^9 {\rm M_{\odot}}$, comparable to the inferred mass 
of the super-massive black-holes at the centres of
M87~\citep{marconi+97} and Cygnus A~\citep{tadhunter+03}.
Thus, the basic physical quantities derived for the quasars in our
sample appear  to be entirely reasonable, requiring neither unorthodox
methods of accretion, nor surprisingly massive black-holes.

It is important to note that the black-hole mass calculation applied
above (and hence the values of $M_{BH}$ and $L_{N}/L_{Edd}$ given in
table 6 and Fig.6) assume a single, fixed value for the
black-hole:spheroid mass ratio. At some level, this is clearly
unrealistic and it is therefore not in fact obvious to what extent the
scatter in the nuclear:host ratio reflects a range of fuelling
efficiency. Accordingly, we conclude this paper with the first
exploration of the extent to which scatter in nuclear:host ratio can or
cannot be explained by intrinsic scatter in the underlying
black-hole:spheroid mass relationship.

\subsection{Black-hole mass versus fuelling rate}
The nature of the link between quasar luminosity and black-hole mass
is more easily explored by plotting host versus nuclear
luminosity. This is shown in Fig.7 where we have plotted the absolute
magnitudes of the hosts against those of the nuclei in our sample
(circles), with $100\%$, $10\%$ and $1\%$ of the Eddington limit shown
as solid, dashed and dotted lines respectively. 
Shown also are points from \citet{dunlop+03} (diamonds) and
\citet*{mcleod+99} (triangles), converted to rest-frame $V$-band, and
our adopted cosmology.  
We have also included 3 objects from the sample of \citet{percival+01}
(stars), for which archival HST images are now available (0043$+$039,
0316$-$346 and 1216$+$069). It now seems likely that seeing
limitations in this  
ground-based study effectively prevented successful disentanglement
of host and nuclear fluxes, and accurate morphological distinctions.
The replacement images from the HST archive have been analysed in
precisely the same way as has been described for the present sample,
and converted into rest-frame $V$-band. 

\begin{figure}
  \centering
  \includegraphics[width=75mm]{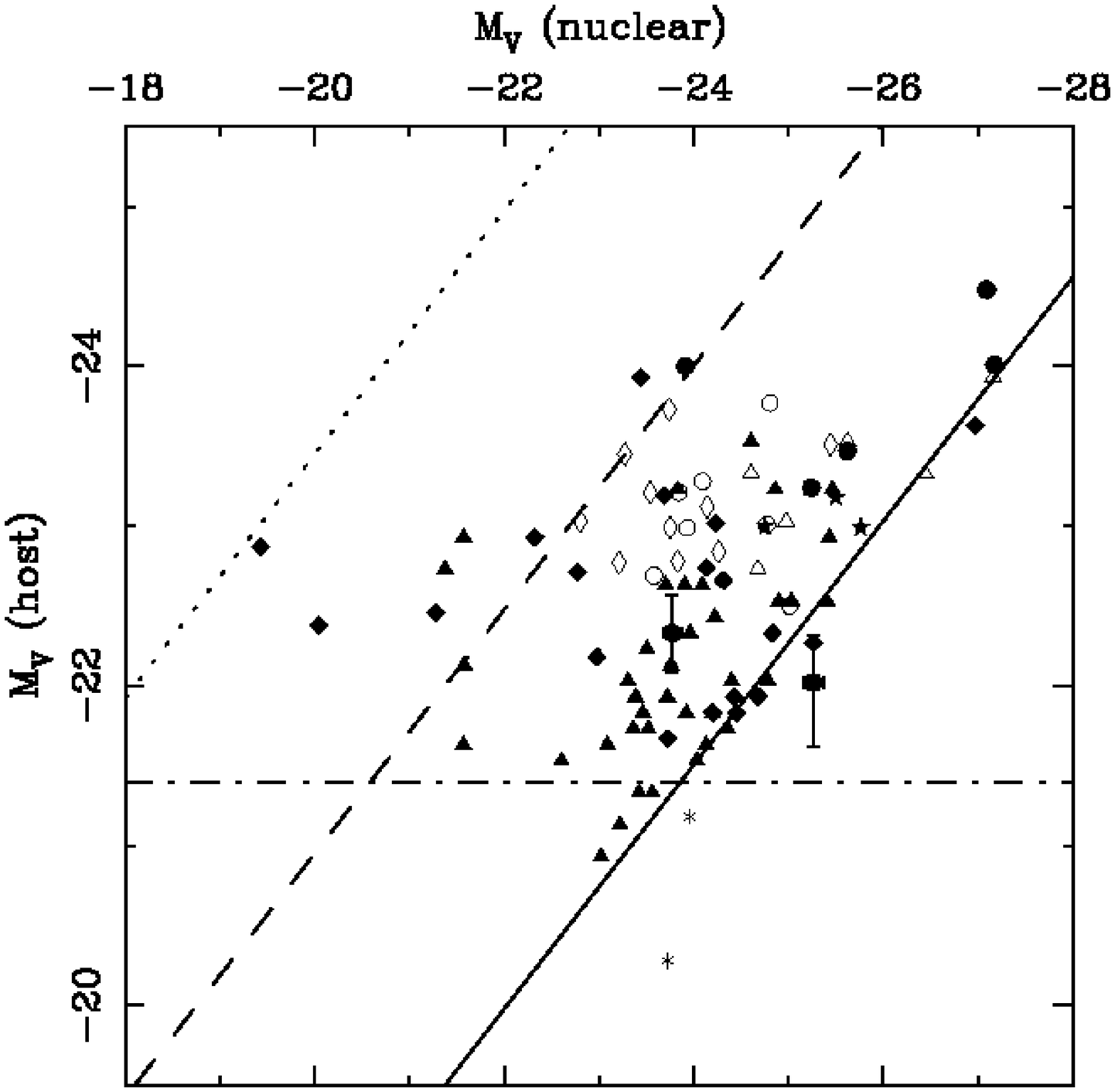}
  \includegraphics[width=75mm]{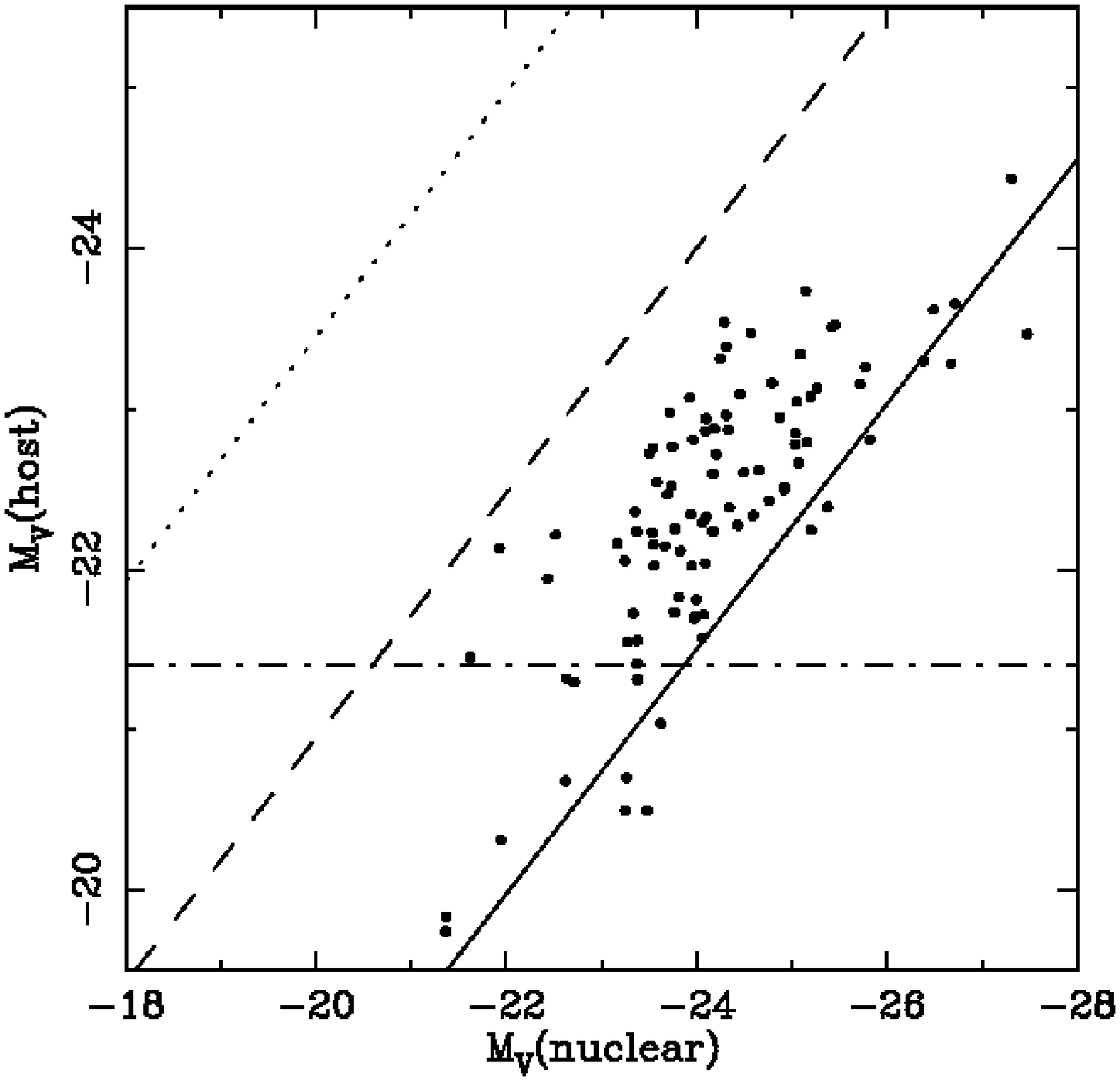}
  \caption{{\bf Upper panel:} Host versus nuclear luminosity for the
    bulge-dominated quasars in the current sample (circles). The
    spheroidal components  of the two disc-dominated quasars,
    1258$-$015 and 1237$-$040 are shown by asterisks.
    The samples of McLeod et al. (1999) (triangles), and Dunlop et
    al. (2003) (diamonds), plus three objects from Percival et
    al. (2001) re-imaged with the HST (stars - see main text)
    are also presented. Filled symbols once again denote radio-quiet
    objects, and open symbols radio-loud objects. The solid, 
    dashed and dotted lines represent objects radiating at 100\%, 10\% and
    1\% of the Eddington luminosity respectively on the assumption
    of a fixed black-hole to bulge mass ratio of
    $M_{BH}=0.0012M_{\mathrm Sph}$.  
    The majority of the quasars in our sample appear to be radiating at
    $>$10\% of their Eddington limit (the single super-Eddington object
    is the poorly constrained 1252$+$020 - see Fig.3.).
    Error bars are smaller than the symbols for all objects
    except for 1252$+$020 and 1258$-$015.
    {\bf Lower panel:}
    The predicted distribution on the host versus nuclear luminosity plane 
    for a sample of quasars radiating at 50\% of
    the Eddington limit, given an adopted scatter of 0.3 dex in the 
    black-hole:spheroid mass relation, and including the effect of 
    the exponential cutoff in the luminosity function above $L^{\star}$
    (dot-dashed line). The random sample has been resampled in order to
    reflect the the same distribution of nuclear luminosities as is displayed
    by the combined data in the upper panel.
    This scenario re-produces much (but not all of) 
    the observed scatter in apparent 
    fuelling efficiency without in fact requiring a range of Eddington ratios.
    However, at the same time an assumed scatter of 0.3 dex or lower is 
    required to avoid too many objects apparently breaching the rather 
    solid Eddington limit displayed by the data.}
\end{figure}

The top panel of Fig.7 shows that, 
while central black-holes appear to accrete with a wide range of
efficiencies, the objects we term quasars are generally produced by
black-holes emitting  at $>10$\% of their Eddington limit, residing in
host galaxies with $L > L^{\star}$. 
However, perhaps the most impressive feature of this plot is that, for
a given host galaxy luminosity, the most luminous nuclear source has a
luminosity essentially exactly as would be predicted from the
Eddington limit corresponding to the mass of the central black-hole as
deduced from the relationship $M_{BH} = 0.0012 M_{\mathrm Sph}$. 
In other words, while the statistical correlation between host-galaxy
and nuclear luminosity within these samples may not be very strong,
the relationship between host-galaxy and {\it maximum} nuclear
luminosity appears extremely tight, and completely consistent with
Eddington limited accretion. 
Such a result has previously been noted by \cite{mcleodrieke95} for
lower-luminosity AGN, but to the best of our knowledge, this is the
first time that it has been demonstrated concretely at high luminosity.
So clean is this relation over
two orders of magnitude in $L_{N}$ that it has the potential 
to constrain the size of the scatter in the underlying black-hole:spheroid
relation for massive galaxies. In turn, such constraints can then illuminate
the extent to which the apparent $~1$ dex scatter in fuelling efficiency
can also be explained by intrinsic scatter in the underlying
black-hole:spheroid mass relationship.

A full exploration of this approach is beyond the scope of the present paper
and is in any case better applied to larger and more statistically complete 
samples. However, we have produced the lower panel of Fig.7 to illustrate 
the extent to which the data can be reproduced by folding in an assumed 
underlying scatter in the black-hole:spheroid mass ratio, while assuming 
a single value for the Eddington ratio.

We generated a random population of spheroidal galaxies, and central
black-holes defined by a Schechter function with $\alpha = 1.25$ and
$log_{10}(\frac{M_{tot}^{*}}{M_{\odot}})=11.5$ (where $M_{tot}^{*}$
denotes the turnover in the distribution of total mass, stellar and
dark matter), and assuming a fixed scatter in the black-hole:spheroid
mass relation. From this galaxy / black-hole population we generate a
sample of quasars, assuming a fixed fuelling efficiency, and
constrained to share the same nuclear luminosity distribution as is
found for the real sample in the top panel of Fig.7. 

Adoption of a scatter larger than 0.3 dex produces significantly more
apparently super-Eddington objects than are observed. Conversely,
adoption of a scatter substantially smaller than 0.3 dex can reproduce
the apparently tight Eddington limit more closely, but underpredicts
the apparent scatter in fuelling efficiency, and cannot therefore
reproduce the tight Eddington limit displayed by the data in the top
panel of Fig.7.

We used a 2D K-S test to compare the simulation with the data. By
marginalising over universal efficiency we find the probability, $p$,
that the quasar sample is consistent with a fixed fuelling efficiency. 
We find that for a 0.3dex or smaller scatter, the quasar sample is
inconsistent with a population in which there is a fixed fuelling
efficiency ($p=0.022$). Increasing the scatter produces an improvement
to acceptable levels at 0.4dex ($p=0.157$) and 0.45dex($p=0.198$), but a
scatter as large as 0.5dex is strongly excluded ($p=0.002$).

The figure presented illustrates the situation for a sample of quasars
radiating at 50\% of the Eddington limit, with an assumed scatter
in the underlying black-hole:spheroid mass relation of 0.3 dex. These values 
were chosen for this illustration as the combination which best reproduces
both the apparently tight Eddington limit, and level of scatter displayed
by the data in the upper panel of Fig.7, adequately describing the
spread in nuclear luminosities at $M_{V}<-23$. However, for dimmer
quasars, we find that a variable fuelling rate is essential in order
to explain the full range in $\frac{L_{N}}{L_{H}}$ observed.
 
In summary, from simulations of the sort described above, the observed
apparent tight upper (Eddington) limit on fuelling efficiency can be
used to set an upper limit of 0.3 dex on the scatter in the underlying 
black-hole:spheroid mass relation, consistent with other recently 
derived values \citep{mcluredunlop02, marconihunt03}. A significant
fraction of the scatter observed in the upper panel of Fig.7 can then
still be attributed to the scatter in the underlying mass
relationship, but some variation in assumed efficiency would still
seem to be required to reproduce the most underluminous objects.

Finally, we can use the tightness of the bound placed by the
Eddington line to place constraints on the hosts of higher redshift
quasars, where direct imaging of the host is not possible. 
A quick inspection of Fig.7 reveals that any quasar brighter than
$M_V=-27$ must be found in, or at least end up within, a spheroidal
galaxy brighter than $M_V = -24$, and must be radiating at a rate
close to its Eddington limit. The turnover in the Schechter function
places a natural limit on the abundance of such large galaxies, and
the data appear to show just such a cutoff at a host luminosity of
$M_{V} \simeq -24.5$.


\section{Summary}
\label{sec-conc}
Through the careful analysis of deep HST images, we have succeeded in 
determining the basic properties of the host galaxies of quasars spanning 
a factor of $\simeq 20$ in luminosity, but within a narrow redshift range at 
$z \simeq 0.4$. The sample under study contains both radio-loud and 
radio-quiet quasars, and includes some of the most luminous quasars known 
in the low-redshift universe. 

Our results confirm and extend the trends uncovered in our previous HST-based 
studies of lower luminosity objects \citep{mclure+99,dunlop+03}.
Specifically we find that the hosts of all the radio-loud quasars, and all 
the radio-quiet quasars with $M_V < -24$ are giant elliptical galaxies, 
with luminosities $L > L^{\star}$, and scale lengths $R_e \simeq 10$ kpc. 
Moreover, the Kormendy relation displayed by these host galaxies is 
indistinguishable from that displayed by nearby, inactive ellipticals.

From the luminosities of their hosts we have estimated the masses of
the black-holes which power the quasars using the relationship
$M_{BH}=0.0012M_{\mathrm Sph}$ and hence, via comparison with the quasar
nuclear luminosities, also the efficiency with which each black-hole is 
emitting relative to the Eddington limit. We find that the order-of-magnitude
increase in nuclear luminosity across our sample can be explained by an
increase in characteristic black-hole mass by a factor $\simeq 3$, 
coupled with a comparable increase in typical black-hole fuelling efficiency. 
However, we find no evidence for super-Eddington accretion, and the
largest inferred black-hole mass in our sample is $M_{BH} \simeq 3 \times 
10^{9} {\rm M_{\odot}}$, comparable to the mass of the black-holes at
the centres of M87 and Cygnus A.

We explore whether intrinsic scatter in the underlying 
$M_{BH}:M_{\mathrm Sph}$ relation (rather than a wide range in
fuelling efficiency) 
can explain the observed scatter 
in the $M_V(host):M_V(nuc)$ plane occupied by quasars. We find that the 
observed tight upper limit on the relation between $M_V(host)$ and
maximum $M_V(nuc)$, (consistent with the Eddington limit inferred from
a single constant of proportionality in the $M_{BH}:M_{\mathrm Sph}$)
constrains
the scatter in the underlying black-hole:spheroid mass relation to be 
0.3 dex or smaller, but that this mass-relation scatter can indeed
explain a substantial fraction of the apparent range in fuelling
efficiency displayed by the quasars, particularly those at
$M_{V}<-23$. The scatter also explains objects with exceptionally high 
nuclear-to-host luminosity ratios without the need for super-Eddington
accretion rates. Finally, our results imply that, due to the cutoff in
the Schechter function, any quasar more luminous than $M_V = -27$ must
be destined to end up in a present-day massive elliptical with $M_V
\simeq -24.5$. 


\section*{Acknowledgements}
Based on observations with the NASA/ESA {\it Hubble Space Telescope},
(program ID's 7447 and 8609) obtained at the Space Telescope Science
Institute, which is operated by The Association of Universities for
Research in Astronomy,
Inc. under NASA contract No. NAS5-26555. This research has made use of
the NASA/IPAC Extragalactic Database (NED) which is operated by the
Jet Propulsion Laboratory, California Institute of Technology, under
contract with NASA. James Dunlop acknowledges the enhanced research time 
afforded by the award of a PPARC Senior Fellowship. DJEF, MJK, RJM \& WJP 
also acknowledge PPARC funding. We thank the anoymous referee for a
fair and constructive critique which has led to significant
improvements in this paper.


\bibliographystyle{mn2e}

\bibliography{floyd04}

\begin{thebibliography}{}

\bibitem[\protect\citeauthoryear{{Abraham}, {Crawford} \& {McHardy}}
  {{Abraham}, {Crawford} \& {McHardy}}{1992}]
  {abraham+92}
  {Abraham} R.~G.,  {Crawford} C.~S.,    {McHardy} I.~M.,  1992, 
  \apj, 401, 474


\bibitem[\protect\citeauthoryear{{Bahcall}, {Kirhakos}, {Saxe} \& {Schneider}}
  {{Bahcall} et~al.}{1997}]
  {bahcall+97}
  {Bahcall} J.~N.,  {Kirhakos} S.,  {Saxe} D.~H.,    {Schneider} D.~P.,  1997,
  \apj, 479, 642

\bibitem[\protect\citeauthoryear{{Bernardi}, et~al.}
  {{Bernardi} et~al.}{2003}]
  {bernardi+03}
  {Bernardi} M., et~al., 
  2003, \aj, 125, 1849
  
\bibitem[\protect\citeauthoryear{{Block} \& {Stockton}}
  {{Block} \& {Stockton}}{1991}]
  {blockstockton91}
  {Block} D.~L.,  {Stockton} A.,  1991, 
  \aj, 102, 1928

\bibitem[\protect\citeauthoryear{{Blundell}, {Beasley}, {Lacy} \& {Garrington}}
  {{Blundell} et~al.}{1996}]
  {blundell+96}
  {Blundell} K.~M.,  {Beasley} A.~J.,  {Lacy} M.,    {Garrington} S.~T.,  1996,
  \apjl, 468, L91

\bibitem[\protect\citeauthoryear{{Blundell} \& {Rawlings}}{{Blundell} \&
    {Rawlings}}{2001}]{blundellrawlings01}
  {Blundell} K.~M.,  {Rawlings} S.,  2001, 
  \apjl, 562, L5

\bibitem[\protect\citeauthoryear{{Boyce}, et~al.}
  {{Boyce} et~al.}{1998}]
  {boyce+98}
  {Boyce} P.~J., et~al.,
  1998, \mnras, 298, 121

\bibitem[\protect\citeauthoryear{{Boyce}, {Disney} \& {Bleaken}}{{Boyce}
    et~al.}{1999}]{boyce+99}
  {Boyce} P.~J.,  {Disney} M.~J.,    {Bleaken} D.~G.,  1999, \mnras, 302, L39

\bibitem[\protect\citeauthoryear{{Dunlop}, {McLure}, {Kukula}, {Baum}, {O'Dea}
  \& {Hughes}}{{Dunlop} et~al.}{2003}]{dunlop+03}
{Dunlop} J.~S.,  {McLure} R.~J.,  {Kukula} M.~J.,  {Baum} S.~A.,  {O'Dea}
  C.~P.,    {Hughes} D.~H.,  2003, \mnras, 340, 1095

\bibitem[\protect\citeauthoryear{{Efstathiou}, {Ellis} \&
  {Peterson}}{{Efstathiou} et~al.}{1988}]{EEP88}
{Efstathiou} G.,  {Ellis} R.~S.,    {Peterson} B.~A.,  1988, \mnras, 232, 431

\bibitem[\protect\citeauthoryear{{Falomo}, {Kotilainen} \& {Treves}}{{Falomo}
  et~al.}{2001}]{falomo+01}
{Falomo} R.,  {Kotilainen} J.,    {Treves} A.,  2001, \apj, 547, 124

\bibitem[\protect\citeauthoryear{{Ferrarese} \& {Merritt}}{{Ferrarese} \&
  {Merritt}}{2000}]{ferraresemerritt00}
{Ferrarese} L.,  {Merritt} D.,  2000, \apjl, 539, L9

\bibitem[\protect\citeauthoryear{{Gebhardt}, et~al.}
  {{Gebhardt} et~al.}{2000}]
  {gebhardt+00}
  {Gebhardt} K.,  et~al.,
  2000, \apjl, 539, L13

\bibitem[\protect\citeauthoryear{{Goldschmidt}, {Kukula}, {Miller} \&
  {Dunlop}}{{Goldschmidt} et~al.}{1999}]{goldschmidt+99}
{Goldschmidt} P.,  {Kukula} M.~J.,  {Miller} L.,    {Dunlop} J.~S.,  1999,
  \apj, 511, 612

\bibitem[\protect\citeauthoryear{{Goldschmidt}, {Miller}, {La Franca} \&
  {Cristiani}}{{Goldschmidt} et~al.}{1992}]{goldschmidt+92}
{Goldschmidt} P.,  {Miller} L.,  {La Franca} F.,    {Cristiani} S.,  1992,
  \mnras, 256, 65P

\bibitem[\protect\citeauthoryear{{Green} \& {Yee}}{{Green} \&
  {Yee}}{1984}]{greenyee84}
{Green} R.~F.,  {Yee} H.~K.~C.,  1984, \apjs, 54, 495

\bibitem[\protect\citeauthoryear{{Gregory}, {Vavasour}, {Scott} \&
  {Condon}}{{Gregory} et~al.}{1994}]{gregory+94}
{Gregory} P.~C.,  {Vavasour} J.~D.,  {Scott} W.~K.,    {Condon} J.~J.,  1994,
  \apjs, 90, 173

\bibitem[\protect\citeauthoryear{{Hamilton}, {Casertano} \&
  {Turnshek}}{{Hamilton} et~al.}{2002}]{hamilton+02}
{Hamilton} T.~S.,  {Casertano} S.,    {Turnshek} D.~A.,  2002, \apj, 576, 61

\bibitem[\protect\citeauthoryear{{Hooper}, {Impey} \& {Foltz}}{{Hooper}
  et~al.}{1997}]{hooper+97}
{Hooper} E.~J.,  {Impey} C.~D.,    {Foltz} C.~B.,  1997, \apjl, 480, L95

\bibitem[\protect\citeauthoryear{{Hutchings}}{{Hutchings}}{1987}]{hutchings87}
{Hutchings} J.~B.,  1987, \apj, 320, 122

\bibitem[\protect\citeauthoryear{{Hutchings}, {Frenette}, {Hanisch}, {Mo},
  {Dumont}, {Redding} \& {Neff}}{{Hutchings} et~al.}{2002}]{hutchings+02}
{Hutchings} J.~B.,  {Frenette} D.,  {Hanisch} R.,  {Mo} J.,  {Dumont} P.~J.,
  {Redding} D.~C.,    {Neff} S.~G.,  2002, \aj, 123, 2936

\bibitem[\protect\citeauthoryear{{Hutchings}, {Johnson} \& {Pyke}}{{Hutchings}
  et~al.}{1988}]{hutchings+88}
{Hutchings} J.~B.,  {Johnson} I.,    {Pyke} R.,  1988, \apjs, 66, 361

\bibitem[\protect\citeauthoryear{{Hutchings} \& {Neff}}{{Hutchings} \&
  {Neff}}{1990}]{hutchings+90}
{Hutchings} J.~B.,  {Neff} S.~G.,  1990, \aj, 99, 1715

\bibitem[\protect\citeauthoryear{{Hutchings} \& {Neff}}{{Hutchings} \&
  {Neff}}{1991}]{hutchingsneff91}
{Hutchings} J.~B.,  {Neff} S.~G.,  1991, \aj, 101, 2001

\bibitem[\protect\citeauthoryear{{Hutchings} \& {Neff}}{{Hutchings} \&
  {Neff}}{1992}]{hutchingsneff92}
{Hutchings} J.~B.,  {Neff} S.~G.,  1992, \aj, 104, 1

\bibitem[\protect\citeauthoryear{{J{\o}rgensen}, {Franx} \&
  {Kjaergaard}}{{J{\o}rgensen} et~al.}{1996}]{jorgensen+96}
{J{\o}rgensen} I.,  {Franx} M.,    {Kjaergaard} P.,  1996, \mnras, 280, 167

\bibitem[\protect\citeauthoryear{{Kormendy} \& {Gebhardt}}{{Kormendy} \&
  {Gebhardt}}{2001}]{kormgeb01}
{Kormendy} J.,  {Gebhardt} K.,  2001, in 20th Texas Symposium on relativistic
  astrophysics {Supermassive Black Holes in Galactic Nuclei (Plenary Talk)}.

\bibitem[\protect\citeauthoryear{{Krist}}{{Krist}}{1999}]{tinytim}
{Krist} J.,  1999, TinyTim User Manual

\bibitem[\protect\citeauthoryear{{Kukula}, {Dunlop}, {McLure}, {Miller},
  {Percival}, {Baum} \& {O'Dea}}{{Kukula} et~al.}{2001}]{kukula+01}
{Kukula} M.~J.,  {Dunlop} J.~S.,  {McLure} R.~J.,  {Miller} L.,  {Percival}
  W.~J.,  {Baum} S.~A.,    {O'Dea} C.~P.,  2001, \mnras, 326, 1533

\bibitem[\protect\citeauthoryear{{Lehnert}, {van Breugel}, {Heckman} \&
  {Miley}}{{Lehnert} et~al.}{1999}]{lehnert+99b}
{Lehnert} M.~D.,  {van Breugel} W.~J.~M.,  {Heckman} T.~M.,    {Miley} G.~K.,
  1999, \apjs, 124, 11

\bibitem[\protect\citeauthoryear{{M{\' a}rquez}, {Petitjean}, {Th{\' e}odore},
  {Bremer}, {Monnet} \& {Beuzit}}{{M{\' a}rquez} et~al.}{2001}]{marquez+01}
{M{\' a}rquez} I.,  {Petitjean} P.,  {Th{\' e}odore} B.,  {Bremer} M.,
  {Monnet} G.,    {Beuzit} J.-L.,  2001, \aap, 371, 97

\bibitem[\protect\citeauthoryear{{Magorrian}, et~al.}
  {{Magorrian} et~al.}{1998}]
  {magorrian+98}
  {Magorrian} J., et~al., 
  1998, \aj, 115, 2285

\bibitem[\protect\citeauthoryear{{Malkan}}{{Malkan}}{1984}]{malkan84}
{Malkan} M.~A.,  1984, \apj, 287, 555

\bibitem[\protect\citeauthoryear{{Marconi}, {Axon}, {Macchetto}, {Capetti},
  {Soarks} \& {Crane}}{{Marconi} et~al.}{1997}]{marconi+97}
{Marconi} A.,  {Axon} D.~J.,  {Macchetto} F.~D.,  {Capetti} A.,  {Soarks}
  W.~B.,    {Crane} P.,  1997, \mnras, 289, L21

\bibitem[\protect\citeauthoryear{{Marconi} \& {Hunt}}{{Marconi} \&
  {Hunt}}{2003}]{marconihunt03}
{Marconi} A.,  {Hunt} L.~K.,  2003, \apjl, 589, L21

\bibitem[\protect\citeauthoryear{{McLeod} \& {Rieke}}{{McLeod} \&
    {Rieke}}{1995}]{mcleodrieke95} 
{McLeod} B.~A.,  {Rieke} G.~H.,  1995, \apj, 441, 96

\bibitem[\protect\citeauthoryear{{McLeod} \& {McLeod}}{{McLeod} \&
  {McLeod}}{2001}]{mcleod01}
{McLeod} K.~K.,  {McLeod} B.~A.,  2001, \apj, 546, 782

\bibitem[\protect\citeauthoryear{{McLeod}, {Rieke} \&
  {Storrie-Lombardi}}{{McLeod} et~al.}{1999}]{mcleod+99}
{McLeod} K.~K.,  {Rieke} G.~H.,    {Storrie-Lombardi} L.~J.,  1999, \apjl, 511,
  L67

\bibitem[\protect\citeauthoryear{{McLure} \& {Dunlop}}{{McLure} \&
  {Dunlop}}{2002}]{mcluredunlop02}
{McLure} R.~J.,  {Dunlop} J.~S.,  2002, \mnras, 331, 795

\bibitem[\protect\citeauthoryear{{McLure}, {Dunlop} \& {Kukula}}
  {{McLure} et~al.}{2000}]
  {mclure+00}
  {McLure} R.~J.,  {Dunlop} J.~S.,    {Kukula} M.~J.,  2000, \mnras, 318, 693

\bibitem[\protect\citeauthoryear{{McLure}, {Kukula}, {Dunlop}, {Baum}, {O'Dea}
  \& {Hughes}}{{McLure} et~al.}{1999}]{mclure+99}
{McLure} R.~J.,  {Kukula} M.~J.,  {Dunlop} J.~S.,  {Baum} S.~A.,  {O'Dea}
  C.~P.,    {Hughes} D.~H.,  1999, \mnras, 308, 377

\bibitem[\protect\citeauthoryear{{Percival}, {Miller}, {McLure} \&
  {Dunlop}}{{Percival} et~al.}{2001}]{percival+01}
{Percival} W.~J.,  {Miller} L.,  {McLure} R.~J.,    {Dunlop} J.~S.,  2001,
  \mnras, 322, 843

\bibitem[\protect\citeauthoryear{{Press}, {Teukolsky}, {Vetterling} \&
  {Flannery}}{{Press} et~al.}{1992}]{numrec}
{Press} W.~H.,  {Teukolsky} S.~A.,  {Vetterling} W.~T.,    {Flannery} B.~P.,
  1992, Numerical recipes in FORTRAN. The art of scientific computing.
Cambridge: University Press, 1992, 2nd ed.

\bibitem[\protect\citeauthoryear{{Puchnarewicz}, et~al.}
  {{Puchnarewicz} et~al.}{1992}]
  {puchnarewicz+92}
  {Puchnarewicz} E.~M.,  et~al., 
  1992, \mnras, 256, 589

\bibitem[\protect\citeauthoryear{{Reimers}, et~al.}
  {{Reimers} et~al.}{1995}]
  {reimers+95}
  {Reimers} D.,  et~al.,
  1995, \aap, 303, 449

\bibitem[\protect\citeauthoryear{{Ridgway}, {Heckman}, {Calzetti} \&
  {Lehnert}}{{Ridgway} et~al.}{2001}]{ridgway+01}
{Ridgway} S.~E.,  {Heckman} T.~M.,  {Calzetti} D.,    {Lehnert} M.,  2001,
  \apj, 550, 122

\bibitem[\protect\citeauthoryear{{S\'{e}rsic}}{{S\'{e}rsic}}{1968}]{sersic68}
{S\'{e}rsic} J.~L.,  1968, in Atlas de Galaxes Australes; Vol. Book; Page 1 {Atlas de Galaxes Australes}.

\bibitem[\protect\citeauthoryear{{Smith}, {Heckman}, {Bothun}, {Romanishin} \&
  {Balick}}{{Smith} et~al.}{1986}]{smith+86}
{Smith} E.~P.,  {Heckman} T.~M.,  {Bothun} G.~D.,  {Romanishin} W.,    {Balick}
  B.,  1986, \apj, 306, 64

\bibitem[\protect\citeauthoryear{{Stockton} \& {Ridgway}}{{Stockton} \&
  {Ridgway}}{2001}]{stocktonridgway01}
{Stockton} A.,  {Ridgway} S.~E.,  2001, \apj, 554, 1012

\bibitem[\protect\citeauthoryear{{Tadhunter}, {Marconi}, {Axon}, K., {Robinson}
  \& {Jackson}}{{Tadhunter} et~al.}{2003}]{tadhunter+03}
{Tadhunter} C.,  {Marconi} A.,  {Axon} D.,  K. W.,  {Robinson} T.~G.,
  {Jackson} N.,  2003, \mnras

\bibitem[\protect\citeauthoryear{{Veron-Cetty} \& {Woltjer}}{{Veron-Cetty} \&
  {Woltjer}}{1990}]{veron90}
{Veron-Cetty} M.~.,  {Woltjer} L.,  1990, \aap, 236, 69

\bibitem[\protect\citeauthoryear{{Veron-Cetty} \& {Veron}}{{Veron-Cetty} \&
  {Veron}}{1993}]{VCV1993}
{Veron-Cetty} M.-P.,  {Veron} P.,  1993, {A Catalogue of quasars and active
  nuclei}.
ESO Scientific Report, Garching: European Southern Observatory (ESO), |c1993,
  6th ed.

\bibitem[\protect\citeauthoryear{{Veron-Cetty} \& {Veron}}{{Veron-Cetty} \&
  {Veron}}{2000}]{VCV2000}
{Veron-Cetty} M.-P.,  {Veron} P.,  2000, {A catalogue of quasars and active
  nuclei}.
A catalogue of quasars and active nuclei / M.-P.~Veron-Cetty and P.~Veron.~
  Garching bei Munchen, Germany : European Southern Observatory,
  c2000.~(Scientific report (European Southern Observatory) ; no.~19)

\bibitem[\protect\citeauthoryear{{Voges}, {Aschenbach}, {Boller}, {Br{\"
  a}uninger}, {Briel} \& {Burkert}}{{Voges} et~al.}{1999}]{voges+99}
{Voges} W.,  {Aschenbach} B.,  {Boller} T.,  {Br{\" a}uninger} H.,  {Briel} U.,
     {Burkert} W.,  1999, \aap, 349, 389

\bibitem[\protect\citeauthoryear{{Wright}, {McHardy} \& {Abraham}}{{Wright}
  et~al.}{1998}]{wright+98}
{Wright} S.~C.,  {McHardy} I.~M.,    {Abraham} R.~G.,  1998, \mnras, 295, 799

\bibitem[\protect\citeauthoryear{{Wyckoff}, {Gehren} \& {Wehinger}}{{Wyckoff}
  et~al.}{1981}]{wyckoff+81}
{Wyckoff} S.,  {Gehren} T.,    {Wehinger} P.~A.,  1981, \apj, 247, 750

\bibitem[\protect\citeauthoryear{{Yee} \& {Green}}{{Yee} \&
  {Green}}{1987}]{yeegreen87}
{Yee} H.~K.~C.,  {Green} R.~F.,  1987, \apj, 319, 28

\end{thebibliography}


\appendix

\section{Images \& Notes on individual objects}
For each quasar we show the final reduced $I$-band (F814W/F791W)
HST image (top left), the best-fit model (pure bulge
or pure disc plus nuclear component) to the quasar image 
(top right), the model host galaxy
after removal of the nucleus (bottom left) and the 
model-subtracted residuals (bottom
right). North and East are marked on the object frame (with the arrow
pointing North). The object and model frames are contoured at the same
surface brightness levels. The level of the lowest contour
is given (in $V$ mag.arcsec$^{-2}$) at the top right of the object frame. 
Successive contour levels are separated by 1 mag.arcsec$^{-2}$.

\subsection{Radio-Quiet Quasars}

\noindent{\bf 0624$+$691} (HS0624$+$6907). One of the brightest
quasars in the sky, this object has been the subject of a
comprehensive multi-wavelength study \citep{reimers+95}, which
classified the host galaxy as
a massive elliptical, fitting a de Vaucouleurs profile with a nominal
scale length of 1.8kpc to their PSF-subtracted image. The quasar
appears to lie in a cluster, with a number of small companion objects
visible in the field.

We find the host to be best fit by a giant elliptical galaxy
($R_{1/2}=10$~kpc), with an extremely strong nuclear component
($L_{N}/L_{H}=18$).  The variable-$\beta$ fit returns a value of
$\beta=0.20$, again consistent with a pure de Vaucouleurs elliptical host.

Due to the extreme luminosity of this quasar the masking for the
diffraction spikes was applied over a much larger area than for the
majority of objects in this study. The host galaxy contribution is
obvious out to a radius of at least 5 arcsec, and we used a fitting
radius of 6 arcsec to ensure that all detectable
host light was used to constrain the model parameters.\\

\noindent{\bf 1001$+$291} (TON0028, PG 1001$+$292, 2MASSi
J1004025$+$285535).  This object was studied in some detail by
\citet*{boyce+99}, who claimed two galactic nuclei; one 1.92 arcsec
(14.6 kpc) to the south-west of the quasar nucleus, and the other 2.30
arcsec (15.9 kpc) to the north-east.  However, this claim appears to
be a result of over-subtraction of the nuclear point source, since
\citet{marquez+01} showed that the host possesses prominent spiral
arms and a bar which crosses the nucleus from north east to south
west, although they were unable to fit a surface-brightness profile.

In the current data we also find spiral arms and a nuclear bar which is
clearly visible in the residual image of this object. However,
we find that the surface light profile of the underlying
smooth component is very well fitted by a de Vaucouleurs law
($R_{1/2}\simeq15$~kpc, $\beta = 0.26$), suggesting a bulge-dominated
host. \\

\noindent{\bf 1230$+$097} (LBQS 1230$+$0947).
We find the host galaxy of this quasar to be an
elliptical with a scale length of $R_{1/2}=6$~kpc. There are a
number of other objects in the same field with possible evidence for a
tidal interaction with the nearest object to the north.

Allowing for a variable value of $\beta$ yields a slightly better fit
with $\beta=0.37$, suggesting a somewhat intermediate morphology.\\

\noindent{\bf 1237$-$040} (EQS B1237$-$0359). This object appears to
be interacting with a companion galaxy to the north, and the residual
image shows that a tidal tail has been induced in the quasar host
itself.  The host is best fitted by a large ($R_{1/2}=4$~kpc) disc
galaxy, with the variable-$\beta$ modelling returns a value of
$\beta=0.96$. The residual image shows some excess nuclear flux
which has not been accounted for by the pure disc fit, and the
2-component modelling reveals a luminous spheroidal component
(Bulge/Disc=0.31) of moderate size ($R_{1/2}=3.8$kpc). The luminosity
of the dominant disc is reduced by 0.2mag.
This is the most luminous of the disc-hosted quasars studied here, and
it is interesting to note that it is also the one with the most
significant detection of a spheroidal host galaxy component.\\

\noindent{\bf 1252$+$020} (EQS B1252$+$020, HE 1252$+$0200).  This
is a radio-quiet \citep{goldschmidt+99}, X-ray detected
\citep{voges+99} quasar with a strong UV excess
\citep{goldschmidt+92}. Of all the objects in the current sample, this quasar
proved to be the hardest for which to achieve an unambiguous model fit
to the host galaxy. Indeed, as is shown in Fig.3, we were unable
to constrain the luminosity of the host to the same extent as
for the other quasars: the $\chi^2$ contours in the $\mu - R$ plane have a
slope slightly steeper than 5. One nearby companion had
to be masked out, along with the diffraction spikes, before modelling
could be carried out. In addition, there is a faint region of
nebulosity directly to the north of the quasar.

Our best fit model has an elliptical host with $R_{1/2}=4$~kpc and the
highest nuclear-to-host ratio in the sample, $L_{N}/L_{H}=19.5$,
although as has been stated, the host and nuclear flux have not been
completely disentangled, and there is a large error associated.
The $\beta$ parameter modelling also favours an
elliptical host, with $\beta=0.22$. 
\\

\noindent{\bf 1254$+$021} (EQS B1254$+$0206). This
radio-quiet
\citep{goldschmidt+99} quasar
shows very smooth extended emission, with only weak
diffraction spikes, and an absence of nearby companions.
We find that the object is best fitted by a large
($R_{1/2} = 14$~kpc) elliptical galaxy and a weak nuclear
component. The variable-$\beta$ fit confirms the host morphology,
returning a value of $\beta=0.24$. No improvement is made with the
addition of a discy component.\\

\noindent{\bf 1258$-$015} (EQS B1255$-$0143, 2MASSi J1258152$-$015918).  A
highly compact object, with an almost stellar appearance
in the HST $I$-band image. However, there is sufficient galaxy light
to model, and we find that the best-fit host is a small elliptical
galaxy with a half-light radius of just $1.5$~kpc. This is confirmed by
the variable-$\beta$ model which returns a value of $\beta=0.26$, with
no appreciable improvement to the fit. Little excess flux remains in
the residual image suggesting that the model has accurately accounted
for all the host galaxy light, and the radial profile is a good fit to
the data throughout. However, the $\chi^{2}$ contours in
$R_{1/2}-\mu_{1/2}$ are slightly too steep to be convinced that we
have completely resolved host from nucleus.
\\

\noindent{\bf 1313$-$014} (Q1313$-$0138, EQS B1313$-$0138, LBQS
1313$-$0138).  
This compact object was only modelled out to 3 arcsec, beyond which
the SNR falls below 1.
The nuclear component of this quasar is relatively
weak, with no prominent diffraction spikes visible in the
image. Hence, despite the small angular size of the host, the model
fit to the galaxy is robust. A spiral feature is visible in the
residual image, with possible evidence for a bar passing through the
nucleus.
We find the host to be best fitted by a disc model with $R_{1/2}\simeq
3$~kpc, and this is supported by the variable-$\beta$ model which
returns a best-fit value of $\beta=0.97$. A low-level spheroidal
component is detected by the 2-component modelling, implying a roughly
Eddington accretion rate, although the
improvement in the fit is marginal.\\

\noindent{\bf 1357$-$024} (EQS B1357$-$0227, 2MASSi
J1400066$-$024131).  There are several fainter objects in the field,
suggesting that 1357$-$024 might lie in a relatively rich cluster
environment.  Diffraction spikes from a nearby bright star are visible
in the southeast quadrant of the image.  However, the quasar itself is
a compact source with no discernible diffraction spikes. All companion
objects, and the majority of the southeast region of the image were
masked out of the fit. In addition, we only modelled out to a radius
of 3 arcsec, where we run into the background.
The modelling software shows a strong preference for a disc-dominated
host, although the variable-$\beta$ model returns the unusual value of
$\beta=1.32$. Most likely this is due to the nearby stellar
diffraction spike leading to a flatter profile (and hence higher
$\beta$). The residual image shows a small amount of excess flux in
the nucleus.
No improvement is obtained with the 2-component modelling. However,
for an Eddington limited nuclear source we would anticipate a bulge
component at $M_{V}\simeq -20.8$, which could easily be concealed in
a compact ($<3kpc$) central bulge component.\\

\noindent{\bf 1821$+$643} (E1821$+$643, IRAS 18216$+$6418, 8C 1821$+$643).
The brightest quasar in the current sample, this object has been
extensively studied at many wavelengths. Extremely luminous in the
infrared and also with a strong X-ray component, this was one of the
first radio-quiet quasars to be studied in detail at radio wavelengths and is
known to contain a small radio jet~\citep{blundellrawlings01}.

Although this quasar appears to lie in a rich field, most of the
surrounding objects are believed to be part of a background cluster at
a redshift $z\simeq0.6$. A previous study resolved the host
galaxy, finding it to be large, featureless and red, but failed to
determine its morphology \citep{hutchingsneff91}.  In addition, the
nucleus itself is unusually red, indicating the presence of large
quantities of dust, though no discrete dust lanes have been observed.
\citet{mcleod01} separated the host and nucleus in their $H$-band NICMOS
imaging study, finding a luminous elliptical galaxy of
magnitude $M_{H} = -26.7$, with a nuclear component with $M_{H} = -29.2$,
(when converted into our cosmology).

Because of the prominent diffraction spikes a larger than usual region
of the image was masked prior to modelling. However, since extended
flux is clearly visible in the image out to a radius of at least 6
arcsec, we therefore used this as our fitting radius. The quasar is
best modelled as a large elliptical host ($R_{1/2}\simeq10$~kpc), with
a strong nuclear component ($L_{N}/L_{H}=11$). The variable-$\beta$ model
is good accord with this decision ($\beta=0.22$), with a significant
improvement in the fit. 

The residuals accentuate the nebulous artifact some 4 arcsec
east-southeast of the nucleus, and also appear to show a spiral-like
feature wrapping around to the northeast. Due to the PSF sampling
problems encountered in this highly luminous object, it remains
unclear as to whether the latter is a genuine feature of the host
galaxy, or simply a PSF artifact.\\ 

\subsection{Radio-Loud Quasars}
\noindent{\bf 0031$-$707} (MC4, 2MASSi J0034052$-$702552).
A radio-loud quasar \citep{gregory+94} originally identified as a
Magellanic object due to its proximity to the galactic plane.  There
are a number of companion objects, suggesting that the object lies in
fairly rich cluster environment, with the potential for interactions
with nearby objects. The HST image shows a relatively weak nucleus
(the model fit gives a nuclear/host ratio of $L_{N}/L_{H} \simeq
2$). The host is best fit by an elliptical galaxy model with
$R_{1/2}=11kpc$ (the variable-$\beta$ model gives a value of
$\beta=0.26$, with only a slight improvement in the quality of the
fit). \\

\noindent{\bf 0110$+$297} (B2$-$0110$+$29, 4C 29.02, 2MASSi
J0113242$+$295815). \citet{malkan84} attempted to resolve the host
galaxy of this quasar
from the ground but was prevented from doing so by poor seeing. This quasar
appears fairly compact in our HST image, with prominent diffraction
spikes and a number of other objects nearby on the sky, including a
well-resolved spiral galaxy some 4 arcsec to the east. We found the
best fitting host to be a large elliptical galaxy with
$R_{1/2} \simeq 12$~kpc, confirmed by the variable-beta fit, which
returns a best-fit value of $\beta=0.22$. There is a small symmetrical
circumnuclear artifact present in the residual image.\\

\noindent{\bf 0812$+$020} (PKS0812$+$02, 4C $+$02.23)
An early study of this object \citep*{wyckoff+81} measured the extent
of the nebulosity surrounding the quasar and found it to have a
diameter of some 88 kpc. Subsequent work was carried out by
\citet{hutchings+90}, who fitted an elliptical galaxy profile to the
host and obtained a scale length of around $12$~kpc (converted to our
cosmology). They also claimed to find evidence for a tidal
interaction.

This quasar lies in a crowded region of sky, and consequently a great
deal of masking was required
before modelling could be carried out.  However, the
host itself is relatively bright, and the preference is for a large
elliptical galaxy with a scale length of $\simeq 17$~kpc. The
variable-$\beta$ modelling confirms the morphology of the host,
returning a best-fit value of $\beta=0.23$.  We find some residual
nuclear flux, but no strong evidence for any disturbance or
interaction in the host.
A major improvement in the fit is obtained by adding a moderate disc
component ($R_{1/2}=$3.7kpc, Bulge/Disc=8.2). There is a 0.2mag drop
in the luminosity of the spheroidal component, but the nuclear
component is unchanged.\\

\noindent{\bf 1058$+$110} (AO1058$+$11, PKS 1058$+$11C, 4C 10.30).
There are a number of apparent companion objects, and the clustering
amplitude was studied by \citet{yeegreen87} \& \citet{greenyee84}. However,
\citet{blockstockton91} found these objects to be at a different
redshift from the quasar.  \citet{hutchings87} detected extended
nebulosity around this object, but was unable to fit a radial
profile.

Although the active nucleus appears to be relatively weak in our
image, the host and nuclear
components proved quite difficult to separate. However the model did
converge on a large elliptical host, with $R_{1/2} = 13$~kpc. The
variable-$\beta$ fit returns a slightly intermediate value of
$\beta=0.33$. No
signs of major disturbance are visible, although some circumnuclear
flux remains in the residual image. Addition of a low-level disc
(Bulge/Disc=16.0) results in a significant improvement in the fit,
with no change in the properties of the dominant bulge model.\\

\noindent{\bf 1150$+$497} (LB2136, 4C 49.22).  Several previous
attempts have been made to detect the host galaxy of this Optically
Violent Variable (OVV) quasar.  \citet{malkan84}, using the Palomar
1.5m telescope, and seeing-degraded models of elliptical and disc-like
hosts, claimed to find a massive elliptical host of scale length $25$~kpc
(when converted to our adopted cosmology). However, an exponential
disc was found to give a reasonable fit by \citet{hutchings87} \&
\citet{hutchings+88}, after PSF-subtraction, and 1D profile fitting.
Finally \citet*{wright+98}, by assuming an elliptical galaxy model (the
relatively poor sampling in their data meant that no real
morphological classification could be performed), detected a host in
$K$-band, with $M_K=-27.3\pm0.6$.

The object appears to be elongated along a north-south axis in the
current HST image.  There are several fainter objects some 10 arcsec
to the NE which have previously been conjectured to be associated with
the quasar \citep{hutchings+88}. Our image also shows two objects,
roughly 2 arcsec to the north \& north west of the quasar which were
masked out along with the diffraction spikes prior to modelling.

Our modelling procedure shows a strong preference for an elliptical
host galaxy, with the best-fit model having a scale length of
$R_{1/2} = 8$ kpc. However, the variable-$\beta$ modelling returns a
best-fit value of $\beta=0.36$, intermediate between pure bulge and disc
morphologies. The reason for this discrepancy may be apparent in the
residual image of the object which shows several regions of excess
flux to the south of the nucleus.\\

\noindent{\bf 1208$+$322} (B2$-$1208$+$32, 7C 1208$+$3213).  This quasar
was detected as a soft X-ray source by Einstein \citep{puchnarewicz+92}.
Optically, it appears to be a compact object with
a strong nuclear component. We find an underlying elliptical
host with a scale length $R_{1/2} = 6.5$ kpc.
A slight improvement to the quality of the fit
is obtained by allowing $\beta$ to vary freely, giving a best-fit
value of $\beta=0.32$. 

Although there are no obvious signs of interaction,
there do appear to be a number of small, faint companion objects
surrounding the quasar.  This is the only radio-loud object in the current
sample whose accretion efficiency appears to come close to the Eddington limit
($L_{N}/L_{Edd}=0.76$).\\

\noindent{\bf 1233$-$240} (PKS1233$-$24, [HB89] 1232$-$249).
\citet{wyckoff+81} found an extended nebulosity with a diameter of
some $166$~kpc surrounding this quasar. The object was also imaged by
\citet{veron90}, who found an elliptical host with magnitude
$M_V=-22.7$, but were not able to provide a scale length.

Despite the prominent diffraction spikes of the strong nuclear
component, some galaxy light is clearly visible in our image, and
there are also several other objects in the field.  The best-fit host
is an elliptical with a scale length of about 3kpc.
Examination of the radial profile shows some excess flux compared to
the pure elliptical model and the variable-$\beta$ model returns a
value of $\beta=0.56$ suggesting that a significant disc component is
also present. \\

\bsp

\label{lastpage}

\end{document}